\documentclass[journal ]{new-aiaa}

\usepackage[utf8]{inputenc}
\usepackage{textcomp}

\usepackage{graphicx}
\usepackage{amsmath}
\usepackage[version=4]{mhchem}
\usepackage{siunitx}
\usepackage{longtable,tabularx}

\usepackage{subcaption}
\usepackage{algorithm}
\usepackage{algpseudocode}
\usepackage{bm}
\usepackage{calligra}
\usepackage[T1]{fontenc} 
\usepackage{booktabs}

\title{Diffractive Sail H-Reversal Trajectory: Theoretical Feasibility, Design Strategies, and Applications}
\author{
	Jinkai Zhang \footnote{Master Student, School of Astronautics, sy2515109zjk@buaa.edu.cn.}, 
	Shuyue Fu\footnote{Ph.D. Student, Shen Yuan Honors College, School of Astronautics; fushuyue@buaa.edu.cn.},
	Di Wu\footnote{Associate Professor, School of Astronautics, wudi2025@buaa.edu.cn.},
	Lin Cheng\footnote{Associate Professor, School of Astronautics, chenglin5580@buaa.edu.cn.},	
	and Shengping Gong\footnote{Professor, School of Astronautics, gongsp@buaa.edu.cn, Senior Member AIAA (Corresponding Author).}
}
\affil{Beihang University, Beijing, 100191, People's Republic of China}
\affil{State Key Laboratory of High-Efficiency Reusable Aerospace Transportation Technology, Beijing, 102206, People’s Republic of China}

\begin{document}

\maketitle

\begin{abstract}

With the growing threat of near-Earth asteroid, planetary defense serves as a vital shield against catastrophic disasters.
Kinetic impact utilizing an angular momentum reversal (H-reversal) trajectory of a solar sail is a highly advantageous defense approach.
However, traditional reflective sails (RS) are constrained during these maneuvers by attitude-thrust coupling and a degradation of solar radiation pressure utilization at the high cone angles required for transverse acceleration.
To enhance impact performance and simplify control, this paper proposes an H-reversal impact scheme utilizing a Sun-facing diffractive sail (SFDS) under a one-stage diffraction angle ($\theta_{\text{\normalfont d}}$) strategy and a two-stage $\theta_{\text{\normalfont d}}$ strategy.
The feasible parameter spaces for both strategies were mapped using the hodograph method. 
Tailored trajectory design methods were established for both strategies based on the feasibility analysis. 
Apophis impact scenario was considered, and the corresponding trajectories were constructed. 
Simulations demonstrate that the one-stage $\theta_{\text{\normalfont d}}$ SFDS outperforms RS through a $\SI{21}{\percent}$ increase in the impact velocity and a $\SI{35}{\percent}$ decrease in the mission duration.
Furthermore, the proposed two-stage $\theta_{\text{\normalfont d}}$ strategy yields an additional $\SI{9}{\kilo\meter\per\second}$ gain in impact velocity. 
By utilizing SFDS H-reversal trajectories, this research establishes an emergency planetary defense framework characterized by rapid response and high kinetic energy.
\end{abstract}

\section{Introduction}
\lettrine{T}{he} solar system harbors a vast and dynamic population of near-Earth asteroids (NEAs)~\cite{zamora_caballero_solar_2025,lin_displaced_2026}.
Although the majority of these celestial bodies safely transit our planetary neighborhood, the potential impact of a sizable asteroid poses a huge risk to the Earth's biosphere and human civilization~\cite{wang_analysis_2026,farnocchia_impact_2026}. 
Consequently, the development of robust and highly efficient active planetary defense framework has emerged as a paramount priority within the global aerospace community.
This imperative is particularly pronounced for NEAs requiring long-term impact-hazard assessment, such as 99942 Apophis~\cite{schettino_potential_2026} and 101955 Bennu~\cite{farnocchia_bennu_2021}.
Given its substantial mass and non-negligible probability of orbital intersection with Earth, 
developing rapid-response mitigation capabilities is essential~\cite{li_apophis_2026,schettino_potential_2026}.
To mitigate such threats, various active deflection strategies have been proposed, including kinetic impacts~\cite{song_terminal_2025,jiao_optimal_2023,negri_shallow_2024}, nuclear detonations~\cite{wang_analysis_2026,wie_hypervelocity_2013,pitz_conceptual_2014}, as well as low-thrust mechanisms utilizing persistent, long-term forces such as gravity tractors~\cite{ketema_mass-optimized_2022,chu_dynamics_2024} and the Yarkovsky effect~\cite{spitale_asteroid_2002}.
Among these, the kinetic impactor concept is widely recognized as one of the most reliable planetary defense methods due to its high technical feasibility and direct energy transfer~\cite{national2010defending}.
In a kinetic impact mission, the terminal relative velocity of the impactor directly determines the destructive effectiveness on the target body. 
For an equivalent deflection outcome, a higher impact velocity allows for a reduction in the required mass of the impactor. 
However, due to the exponential mass penalty inherent in the Tsiolkovsky rocket equation, traditional chemical propulsion systems face severe engineering bottlenecks in attaining the hypervelocities~\cite{pitz_conceptual_2014}. 
Furthermore, most early studies primarily focused on prograde impact orbit designs, such as that of NASA's Double Asteroid Redirection Test (DART) mission~\cite{sarli_double_2019}.
This "chase" mode causes the terminal relative velocity to be heavily constrained, making it difficult to achieve the hyper-velocity required for emergency planetary defense.

To break through this bottleneck, solar sails offer a promising solution by facilitating direct access to retrograde impact orbits. 
Solar sails can be classified as reflective sail (RS), refractive sail~\cite{bassetto_optimal_2021}, and diffractive sail~\cite{quarta_optimal_2024,bassetto_diffractive_2024}, with earlier studies focusing primarily on the RS~\cite{macdonald_solar_2011,macdonald_advances_2014,gong_review_2019}.
McInnes~\cite{mcinnes_deflection_2004,mcinnes_high_energy_2005} explored the use of an orbital inclination "cranking" maneuver to transition an RS into a retrograde heliocentric orbit, elevating the terminal impact velocity to at least 60~km/s.
Building upon this framework, Wei~\cite{wie_solar_2005} proposed a multi-impactor deflection architecture utilizing simplified control laws to coordinate sequential kinetic impacts.
Additionally, Dachwald et al.~\cite{dachwald_solar_2007} introduced evolutionary neurocontrol methodologies to further optimize the orbit designs for these high-energy impact missions.
Nevertheless, as demonstrated by Macdonald et al.~\cite{macdonald_solar_2006}, due to the limited propulsive performance of the RS, such inclination cranking maneuvers remain highly time-consuming. 
To break this limitation, Vulpetti~\cite{vulpetti_sailcraft_1997,vulpetti_general_1999,vulpetti_reaching_2011} pioneered the concept of the angular momentum reversal (H-reversal) trajectory for high-performance RS, enabling rapid access to retrograde orbits without the need for complex inclination cranking maneuvers.
Building on this foundational concept, Mengali et al.~\cite{mengali_h2_reversal_2011} introduced two-dimensional (2D) double H-reversal periodic orbits and identified their quasi-heliostationary properties near aphelion. 
Zeng et al.~\cite{zeng_three_dimensional_2011,zeng_new_2011,zeng_solar_2019} subsequently extended these trajectories to three dimensions, and conducted preliminary investigations into utilizing H-reversal trajectories for asteroid deflection scenarios~\cite{zeng_new_2011}.
Subsequently, Gong et al.~\cite{gong_utilization_2011} performed an in-depth trajectory optimization for an RS impacting Apophis, proposing a piecewise attitude control strategy,
their strategy achieved an impact velocity of 90~km/s within a flight time of approximately 600~days.

However, utilizing an RS to achieve an H-reversal orbit presents inherent limitations. 
To obtain the maximum transverse thrust, RS must maintain a highly inclined cone angle relative to the incident sunlight. 
This inclination not only causes a reduction of solar radiation pressure (SRP) utilization, 
but also induces a strong nonlinear coupling between the thrust direction and the spacecraft's attitude. 
Furthermore, the relatively low rate of angular momentum change results in long transfer times, making it difficult to meet the rapid-response requirements of emergency planetary defense missions.
To address the limitations of RS, Sun-facing diffractive sail (SFDS) technology offers a solution. 
SFDS alters photon momentum utilizing diffraction grating structures. This enables the SFDS to passively maintain a sun-facing attitude while generating stable transverse thrust~\cite{chu_minimum-time_2024}. 
In recent years, extensive research has been dedicated to evaluating SFDS capabilities across various mission scenarios, including heliocentric time-optimal transfers and rendezvous missions~\cite{quarta_optimal_2023,chu_minimum-time_2024}, solar polar imagers~\cite{chu_enhanced_2024}, and Earth-polar observation architectures~\cite{chu_potential_2024}.
Applying SFDS to H-reversal missions eliminates energy loss from large cone angles and decouples thrust direction from attitude, potentially shortening mission times and exceeding the impact velocity limits compared to RS.
Several models have been developed to characterize the diffractive radiation pressure acting on an SFDS, 
including the ideal diffraction-grating model~\cite{quarta_optimal_2023}, 
the electro-optically controlled panels (EOCPs) model~\cite{quarta_solar_2023}, 
the liquid-crystal optical phased-array (LC-OPA) model~\cite{chu_controllable_2021}, 
and the liquid-crystal polarization-gratings (LCPGs) model~\cite{yao_orbit_2024}. 
Among these, the ideal diffraction-grating model provides a more general and well-established theoretical framework.
In addition to the selection of the model, feasibility analysis of the SFDS (also including the RS) H-reversal trajectories remains an open problem.
Wokes et al.~\cite{wokes_classification_2008} provided a classification of 2D fixed attitude RS trajectories under specific parameter combinations.
Zeng et al.~\cite{zeng_feasibility_2011} established a procedure that deduced the necessary conditions for RS H-reversal domains. 
In this paper, however, we can further explore the sufficient and necessary conditions for the H-reversal mechanism based on the dynamical analysis.

In light of these considerations, 
this paper establishes a research framework for SFDS H-reversal trajectories to achieve rapid-response and hyper-velocity asteroid impact mission, including theoretical parameter feasibility domains, tailored design strategies, and planetary impact application.
Based on Wokes~\cite{wokes_classification_2008} and Zeng et al.~\cite{zeng_feasibility_2011}, the necessary and sufficient conditions for generating SFDS 2D H-reversal trajectories under one-stage $\theta_{\text{d}}$ strategy are established.
To further enhance impact performance while balancing engineering feasibility and model universality, a two-stage $\theta_{\text{d}}$ strategy is proposed.
The parameter feasibility domains and boundary characteristics subject to perihelion distance constraints under both strategies are derived.
Regarding the constrained trajectories design on asteroid impact missions, a Perihelion-Constrained Phase-Matching Algorithm is developed for the one-stage $\theta_{\text{\normalfont d}}$ strategy.
Based on the feasibility analysis and the results of one-stage $\theta_{\text{\normalfont d}}$ strategy, a tailored optimization problem is established for the two-stage $\theta_{\text{\normalfont d}}$ strategy.
The procedure could also be extended to RS.
Finally, taking the Asteroid 99942 Apophis impact mission as a representative scenario, numerical simulations are conducted to evaluate and compare the performance of both strategies against the conventional RS. 
The results demonstrate that under the constraints of $\beta = 0.7$ and $r_{\text{p,target}} = 0.3\textnormal{ AU}$,
the one-stage $\theta_{\text{\normalfont d}}$ SFDS strategy increases the impact velocity by 21\%
and shortens the mission time by 35\% compared with RS.
Furthermore, the proposed two-stage $\theta_{\text{\normalfont d}}$ control strategy yields an additional velocity increment of $\SI{9}{\kilo\meter\per\second}$. 

The remainder of this paper is structured as follows: 
Section II develops the dynamical model of the SFDS, 
while Section III maps the feasibility domains of both one-stage $\theta_{\text{d}}$ and two-stage $\theta_{\text{d}}$ strategies for the 2D H-reversal trajectories. 
Section IV details the H-reversal trajectory design methods for asteroid impact mission.
These methodologies are subsequently applied to the Asteroid Apophis mission scenario in Section V, accompanied by comprehensive simulation analyses and comparison. 
Finally, Section VI concludes the study.

\section{Dynamical Models}
\label{sec:Dynamical Models}
This section formulates the 2D heliocentric dynamical model and the ideal force model for SFDS, providing a foundation for analyzing the H-reversal trajectory.

\subsection{Equations of Motion and SRP Models}
To describe the planar motion, a heliocentric polar coordinate system $\{r, \theta\}$ is established with its origin at the solar center of mass. The polar angle $\theta$ is measured from the vernal equinox, which serves as the fixed inertial reference direction. The state vector is denoted by $\bm{X}=[\bm{r}^{\top},\bm{v}^{\top}]^{\top}=[r,\theta,v_r,v_{\theta}]^\top$.
The SFDS's planar motion is governed by the following equations:
\begin{equation}
	\dot{r} = v_r,
	\qquad
	\dot{\theta} = \frac{v_{\theta}}{r},
	\qquad
	\dot{v_r} = \frac{v_{\theta}^2}{r} - \frac{\mu}{r^2} + a_r,
	\qquad
	\dot{v_\theta} = -\frac{v_r v_{\theta}}{r} + a_{\theta},
	\label{eq:EOM}
\end{equation}
where $\mu$ denotes the gravitational parameter of the Sun. $a_r$ and $a_{\theta}$ represent the radial and transverse components of the SRP acceleration, respectively. 
For numerical computations, all variables are normalized. The distance is scaled by the astronomical unit $1\text{ LU} = 1\text{ AU}$, and gravitational parameter is set to $\mu = 1$. Accordingly, the unit of time is derived as $1\text{ TU} = \sqrt{\text{LU}^3/\mu} \approx 58.1310\text{ days}$. 

\begin{figure}[htbp]
	\centering
	\includegraphics[width=0.25\columnwidth]{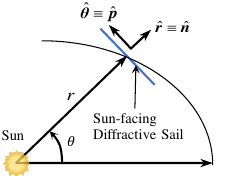}
	\caption{Schematic of coordinate.}
	\label{fig:coordinate}
\end{figure}

\begin{figure*}[htbp]
	\centering
	\begin{subfigure}{0.29\textwidth}
			\centering
			\includegraphics[height=5.0cm]{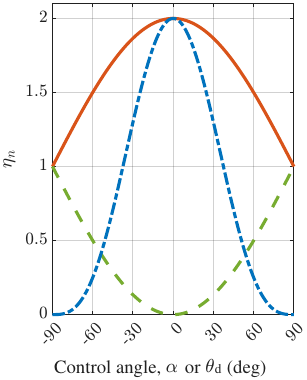}
			\caption{Normal}
			\label{fig:force_radial}
		\end{subfigure}
	\hfill
	\begin{subfigure}{0.29\textwidth}
			\centering
			\includegraphics[height=5.0cm]{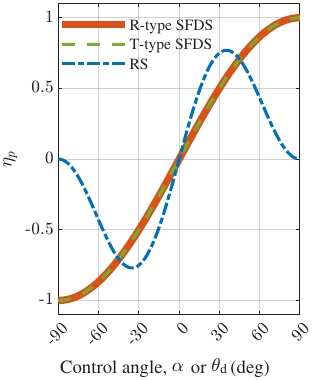}
			\caption{Transverse}
			\label{fig:force_transverse}
		\end{subfigure}
	\hfill
	\begin{subfigure}{0.29\textwidth}
			\centering
			\includegraphics[height=5.0cm]{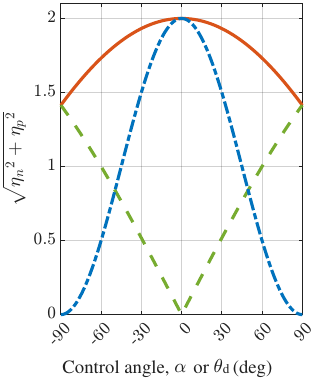}
			\caption{Total}
			\label{fig:force_total}
		\end{subfigure}
	\caption{Comparison of dimensionless force efficiency factors versus control angles across R-type SFDS, T-type SFDS, and RS.}
	\label{fig:force_comparison}
\end{figure*}

A SFDS utilizes microstructured surface gratings to redirect incident photons and assumed to maintain a strict Sun-facing configuration, that is the normal of the sail is always aligned with the incident sunlight.
\begin{figure*}[htbp]
	\centering
	\begin{subfigure}{0.36\textwidth}
		\centering
		\includegraphics[width=\linewidth]{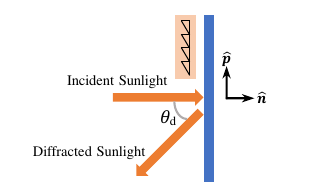}
		\caption{R-type SFDS}
		\label{fig:ds_sub_a}
	\end{subfigure}
	\begin{subfigure}{0.36\textwidth}
		\centering
		\includegraphics[width=\linewidth]{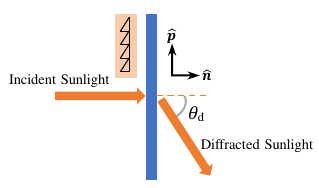}
		\caption{T-type SFDS}
		\label{fig:ds_sub_b}
	\end{subfigure}
	\caption{Schematic of solar radiation pressure force generation for R-type and T-type SFDS (adapted from~\cite{chu_minimum-time_2024}).}
	\label{fig:ds_schematic}
\end{figure*}
The transverse force is generated via the inherent diffraction angle within the microstructures rather than mechanical tilting.
As illustrated in Fig.~\ref{fig:ds_schematic}, the SFDS is categorized into reflection-type (R-type) and transmission-type (T-type) configurations~\cite{chu_dynamics_2024}, for simplicity, we use ideal models.
The SRP-induced acceleration components along the radial and transverse directions are formulated using the normal and transverse force efficiency factors, denoted as $\eta_n$ and $\eta_p$, respectively:
\begin{equation}
	a_r = \frac{\beta}{2r^2} \eta_n,
	\qquad
	a_{\theta} = \frac{\beta}{2r^2} \eta_p,
\end{equation}
$\beta=1.53/\sigma$ is lightness number, $\sigma$ is the density of the sail, which unit is $\SI{}{\gram\per\square\meter}$.
In this study, we restrict $\beta$ to the range $\beta < 1$.
These efficiency factors depend on the microscopic grating type and are formulated as:
\begin{equation}
	\eta_n = 1 \pm \cos \theta_{\text{d}},
	\qquad
	\eta_p = \sin \theta_{\text{d}},
	\qquad
	\theta_{\text{d}} \in \left[ -\frac{\pi}{2},\frac{\pi}{2} \right],
\end{equation}
where the plus sign ($+$) for R-type, and the minus sign ($-$) for T-type. 
Furthermore, to facilitate a comprehensive comparative analysis with the RS, the efficiency factors of the RS are also presented as functions of its attitude angle $\alpha$:
$\eta_{n} = 2 \cos^3\alpha $, $ \eta_{p} = 2 \cos^2\alpha \sin\alpha$.
where $\alpha$ represents the cone angle, defined as the angle between the sail surface normal and the incident sunlight.
To intuitively contrast the thrust generation capabilities and inherent physical limitations of these sail architectures, 
Fig.~\ref{fig:force_comparison} depicts the dimensionless normal, transverse, and total force efficiency factors as functions of their primary control angles.
For H-reversal trajectories, the evolution of specific angular momentum $h = rv_{\theta}$ is governed by:
\begin{equation}
	\dot{h} = \dot{r}v_\theta + r\dot{v_\theta}=v_r v_\theta+r( -\frac{v_r v_{\theta}}{r} + a_{\theta}) = r a_{\theta}.
\end{equation}
To ensure $\dot{h} < 0$, a negative transverse acceleration $a_{\theta} < 0$ is required, which dictates a sail configuration range of $\theta_{\text d} \in [-\pi/2, 0)$ (similarly, $\alpha \in [-\pi/2, 0)$ for RS) . This requirement is uniformly applied to both R-type and T-type SFDS throughout the following analysis.

\subsection{One-Stage $\theta_{\text{\normalfont d}}$ and Two-Stage $\theta_{\text{\normalfont d}}$ Strategies}
The one-stage $\theta_{\text{d}}$ strategy assumes that the $\theta_{\text{d}}$ remains constant throughout the entire mission:
\begin{equation}
	\theta_{\text{d}}(t) = \theta_{\text{d,one-stage}}.
	\label{eq:fixed}
\end{equation}
Although structurally straightforward, the one-stage $\theta_{\text{d}}$ strategy restricts the trajectory modulation capability, making it challenging to satisfy constraints or fully exploit the SFDS's performance potential.

To enhance the orbital control capability while maintaining model simplicity for universal dynamics analysis, this paper proposes a discrete two-stage $\theta_{\text{d}}$ control strategy. 
The engineering feasibility 
is supported by recent advancements in micro-nano optics and non-mechanical beam steering technologies. 
Specifically, Gong et al.~\cite{gong_dynamic_2025} demonstrated that a cascaded multi-layer LCPGs architecture can generate a set of discrete, controllable $\theta_{\text{d}}$ values with a switching time on the order of milliseconds. 
Wang~\cite{wang_reflective_2025} utilized a reflective LCPGs configuration to achieve large $\theta_{\text{d}}$ that is approximately twice that of a transmissive type. 
However, increasing the number of $\theta_{\text{d}}$ requires more LCPGs layers, which escalates manufacturing complexity, increases the areal density, and induces Fresnel losses at the layers interfaces~\cite{kim_wide_angle_2008}. 
Therefore, the number of switchable $\theta_{\text{d}}$ should preferably be kept to a minimum. 
This practical trade-off justifies the selection of a discrete two-stage $\theta_{\text{d}}$ strategy rather than a multi-stage or continuous $\theta_{\text{d}}$ alternative.
Within this framework, the $\theta_{\text{d}}$ is allowed to switch between two pre-designed discrete values, determined by the surface grating microstructures. 
To simplify control and account for the characteristics of H-reversal trajectories (detailed in Section~\ref{subsec:two_stage_feasibility}), the $\theta_{\text{d}}$ is switched only once.
The strategy can be formulated as:
\begin{equation}
	\theta_{\text{d}}(t) = 
	\begin{cases} 
		\theta_{\text{d}1}, & t_0 \le t < t_{\text{sw}}, \\ 
		\theta_{\text{d}2}, & t_{\text{sw}} \le t \le t_{\text{f}}, 
	\end{cases}
	\label{eq:two_stage}
\end{equation}
where $t_0$, $t_{\text{sw}}$, and $t_{\text{f}}$ denote the initial epoch, the switching epoch, and the final mission time, respectively. 
Each stage operates under the ideal SFDS force model described in the preceding subsection, 
and the response time during the transition is assumed to be negligible.

\section{Feasibility Analysis of 2D H-reversal trajectory}
\label{sec:feasibility_analysis}
This section evaluates the feasible parameter space required to achieve the SFDS 2D H-reversal trajectory under both one-stage $\theta_{\text{d}}$ and two-stage $\theta_{\text{d}}$ strategies. 
For simplicity, this section assumes that Earth's orbit is circular with a radius of $1~\text{AU}$ and models the Sun as a point mass. Upon departing Earth's sphere of influence at the departure epoch $t_0 = 0$ (where $\theta_0 = 0^\circ$), the initial heliocentric state of the sailcraft matches that of Earth, i.e., $\bm{X}_0 = \bm{X}_{\text{earth}}(t_0) = [1, 0, 0, 1]^\top$.

\subsection{Hodograph Method}
By mapping the equations of motion into the velocity space, the hodograph method effectively reduces the dimension of the dynamical system. 
To parameterize the SRP components, two dimensionless parameters, $\eta$ and $\xi$, are introduced to reparameterize the radial and transverse accelerations:
\begin{equation}
	{a_r} = \frac{\mu }{{{r^2}}}\left( {\eta  + 1} \right) = \frac{\mu }{{2{r^2}}}{\beta \left( {1 \pm \cos {\theta _{\text{d}}}} \right)},
	\qquad 
	{a_{\theta}} =  - \frac{\mu }{{{r^2}}}\xi \eta = \frac{\mu }{{2{r^2}}}{\beta \sin {\theta _{\text{d}}}}.
	\label{eq:xi_eta_relation_with_beta_thetad}
\end{equation}
This study restricts the lightness number $\beta < 1$, which implies $\eta < 0$. 
To satisfy the continuous deceleration requirement of $a_\theta < 0$ for an H-reversal trajectory, the parameter $\xi$ must also satisfy $\xi < 0$. 
By introducing the transformed dimensionless velocity variables
\begin{equation}
	v = \frac{{h{v_\theta}}}{\mu } = \frac{{{h^2}}}{{\mu r}},
	\qquad 
	w = \frac{{h{v_r}}}{\mu } = \frac{{h\dot r}}{\mu }.
\end{equation}
Eqs.~\eqref{eq:EOM} can be transformed into the following compact form:
\begin{equation}
	v' =  - w - 2\eta \xi ,
	\qquad 
	w' = v - \frac{{\eta \xi w}}{v} + \eta .
	\label{eq:phase_space}
\end{equation}
Equations.~\eqref{eq:phase_space} are differentiated with respect to $\theta$ rather than $t$ (i.e., $(\cdot)' = \frac{\text{d}}{\text{d}\theta}$).
Based on the relation $\dot{\theta} = h/r^2$, 
the behavior in physical space depends on the sign of $h$: for $h > 0$, the physical evolution corresponds to forward integration along the trajectories in $v-w$ space, whereas for $h < 0$, the physical evolution corresponds to backward integration in $v-w$ space.	
\begin{figure*}[htbp]
	\centering
	\begin{subfigure}{0.4\textwidth}
		\centering
		\includegraphics[height=6.0cm]{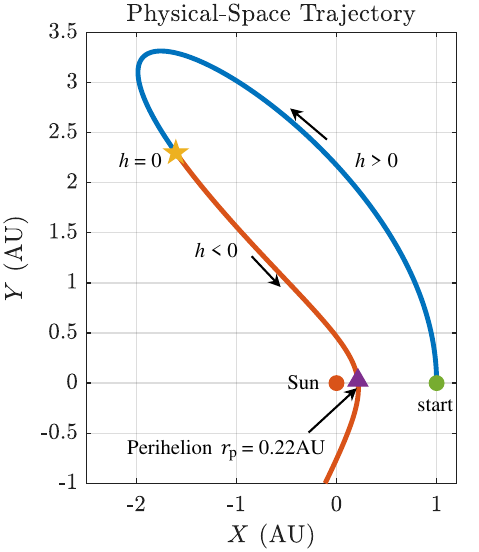}
		\caption{Physical space}
		\label{fig:physical_space}
	\end{subfigure}
	\begin{subfigure}{0.4\textwidth}
		\centering
		\includegraphics[height=6.2cm]{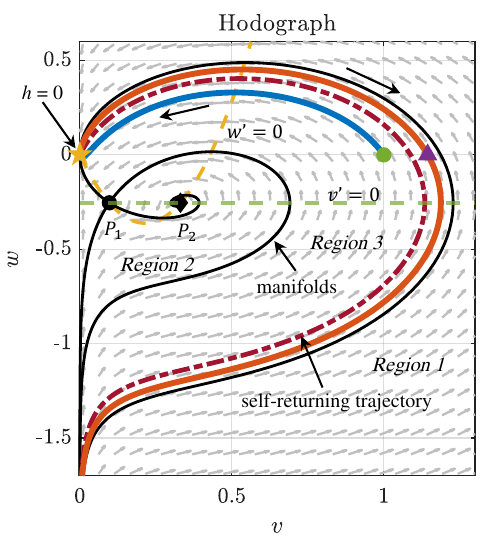}
		\caption{$v-w$ space}
		\label{fig:vw_space}
	\end{subfigure}
	\caption{Physical-space and corresponding $v-w$ space for SFDS 2D H-reversal trajectory $\beta = 0.6 $ $\theta_{\text{\normalfont d}} = -25^{\circ} $.}
	\label{fig:phase_space}
\end{figure*}
As established in previous literature~\cite{wokes_classification_2008}, the topological structure of the $v-w$ phase space depends on the existence of equilibrium points, which is governed by the parameter $\xi$. 
When $|\xi| \le \frac{1}{2\sqrt{2}}$, There are two distinct equilibrium points along the line $w = -2\eta\xi$: a saddle $P_1(-\frac{\eta}{2}\left(1-\sqrt{1-8\xi^2}\right),~-2\eta\xi)$ and a sink $P_2(-\frac{\eta}{2}\left(1+\sqrt{1-8\xi^2}\right), ~-2\eta\xi)$.
As shown in Fig.~\ref{fig:vw_space},
The $v-w$ space is divided into three distinct dynamical regions by the manifolds originating from $P_1$ and $P_2$.
Initial states belonging to different regions lead to different asymptotic behaviors. 
For clarity, regions in the $v\text{-}w$ space are denoted in italics, whereas those in the subsequent parameter space are presented in roman font. 
The characteristics of these three regions in the $v\text{-}w$ space are detailed below:
\begin{enumerate}
	\item \textit{Region 1} (Direct Escape Region): Trajectories integrated backward in $\theta$ lead to solar escape with $h > 0$. Forward integration passes through the origin $(0, ~0)$ where $h$ changes sign, and subsequently results in escape with $h < 0$. The radial velocity $\dot{r}$ changes sign exactly once.
	\item \textit{Region 2} (Spiral-Inward Region): Backward integration in $\theta$ leads to solar escape, whereas forward integration results in the sailcraft falling into the Sun. 
	The sign of $h$ remains unchanged.
	\item \textit{Region 3} (H-Reversal Region): Similar to \textit{Region 1}, except that $\dot{r}$ changes sign three times, which enables the generation of H-reversal trajectories. Notably, this domain features a specific trajectory known as the self-returning trajectory, where $\dot{r} = 0$ occurs precisely at the instant when $h = 0$. Initial states located below and above this trajectory yield internal and external H-reversal trajectories, respectively.
\end{enumerate}
Besides, when $|\xi| > \frac{1}{2\sqrt{2}}$, no equilibrium points exist (which corresponds to the Region 4 of the parameter space discussed below), and H-reversal trajectories still remain feasible. 
Figure~\ref{fig:physical_space} shows a typical H-reversal trajectory of the SFDS, while Fig.~\ref{fig:vw_space} illustrates its corresponding phase space structure.
Departing from Earth's orbit (corresponding to $(1,0)$ in the $v$--$w$ space) and propelled by a continuous negative transverse diffractive force, the trajectory smoothly passes through the angular momentum reversal point, transitions into a retrograde orbit, and crosses the perihelion ($w=0$), thereby completing the Solar Photonic Assist (SPA) phase.

\subsection{One-Stage $\theta_{\text{\normalfont d}}$ Strategy Feasibility Analysis}

\subsubsection{Unconstrained Perihelion Scenario}
For a given lightness number $\beta$, varying the diffraction angle $\theta_{\text{d}}$ alters the parameters $\xi$ and $\eta$, reshaping the $v-w$ space topology, while the initial state $(1,~0)$ remains stationary.
Consequently, the trajectory behavior can be determined by evaluating the location of the initial state relative to the manifolds. 
When equilibrium points exist, the final state of the trajectory is obtained by identifying which domain (\textit{Region 1}, \textit{2}, or \textit{3}) contains the initial point, as well as its position relative to the self-returning trajectory. 
When no equilibrium points exist, the classification into an internal or external H-reversal trajectory depends solely on the relative positional between the initial point and the self-returning trajectory.
For comparison, we also analyzed the feasible regions of the conventional RS under fixed cone angle strategy, following the same procedure described above.
By calculating over a high-resolution grid across $\beta$ and $\theta_{\text{d}}$ (or $\alpha$), six distinct parameter regions are obtained in each sail configuration.
\begin{figure*}[htbp]
	\centering
	\begin{subfigure}[b]{0.32\textwidth}
		\centering
		\includegraphics[height=5.5cm]{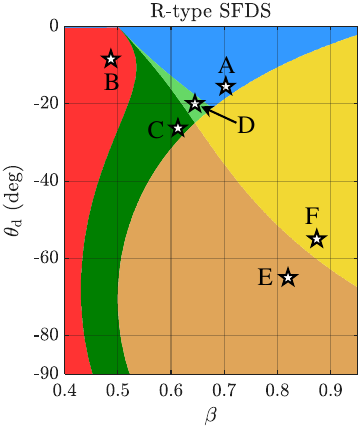}
		\caption{R-type SFDS}
		\label{fig:feasibility_r_type}
	\end{subfigure}
	\hfill
	\begin{subfigure}[b]{0.32\textwidth}
		\centering
		\includegraphics[height=5.5cm]{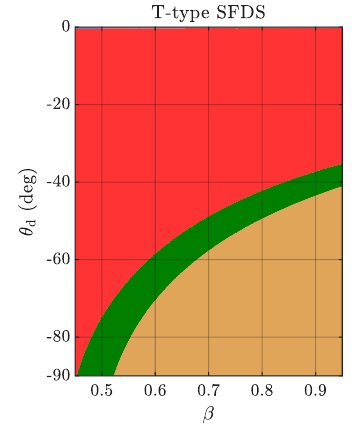}
		\caption{T-type SFDS}
		\label{fig:feasibility_t_type}
	\end{subfigure}
	\hfill
	\begin{subfigure}[b]{0.32\textwidth}
		\centering
		\includegraphics[height=5.5cm]{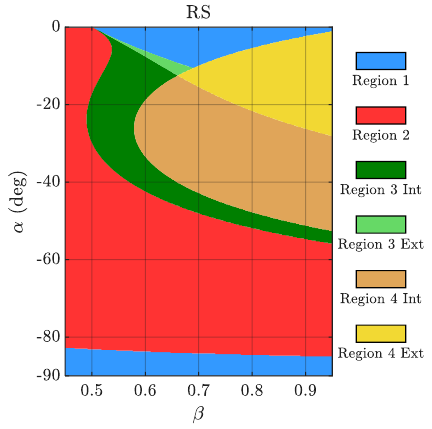}
		\caption{RS}
		\label{fig:feasibility_rs}
	\end{subfigure}
	
	\caption{One-stage $\theta_{\text{\normalfont d}}$ feasibility regions for different sail configurations.}
	\label{fig:feasibility_analysis}
\end{figure*}

\begin{figure*}[htbp]
	\centering
	\captionsetup[subfigure]{font=small,skip=2pt}
	
	\begin{subfigure}[t]{0.45\textwidth}
		\centering
		\includegraphics[width=\linewidth]{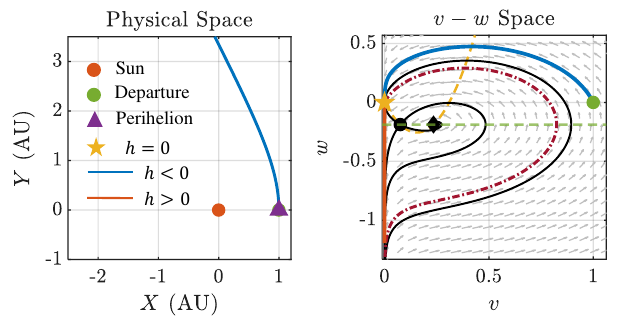}
		\caption*{\(A\quad \beta=0.703\quad \theta_{\mathrm{d}}=-15.5^\circ\)}
	\end{subfigure}
	\begin{subfigure}[t]{0.45\textwidth}
		\centering
		\includegraphics[width=\linewidth]{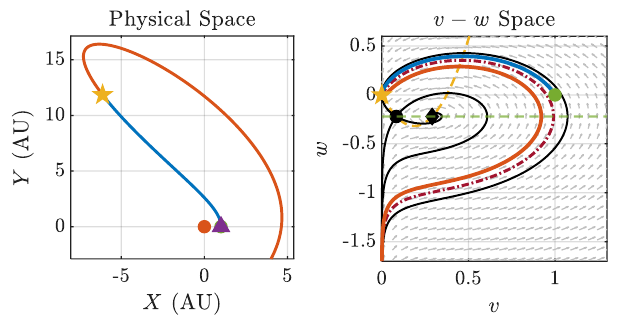}
		\caption*{\(D\quad \beta=0.645\quad \theta_{\mathrm{d}}=-20.0^\circ\)}
	\end{subfigure}
	
	\medskip
	
	\begin{subfigure}[t]{0.45\textwidth}
		\centering
		\includegraphics[width=\linewidth]{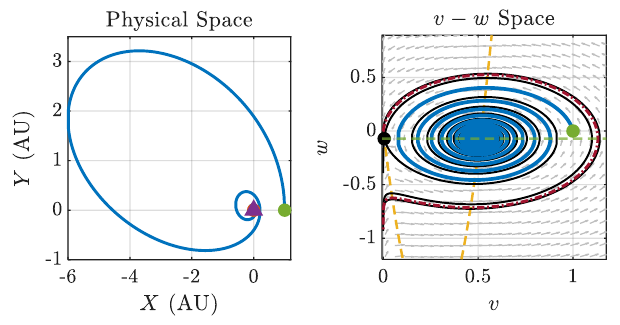}
		\caption*{\(B\quad \beta=0.487\quad \theta_{\mathrm{d}}=-8.42^\circ\)}
	\end{subfigure}
	\begin{subfigure}[t]{0.45\textwidth}
		\centering
		\includegraphics[width=\linewidth]{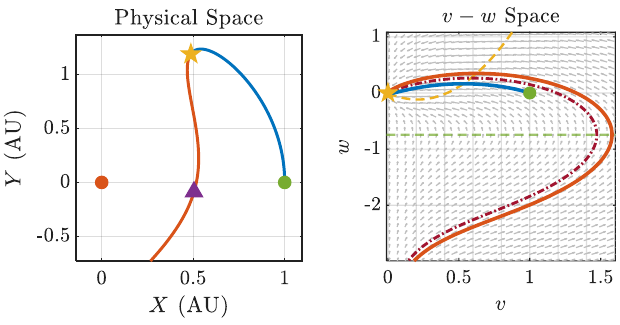}
		\caption*{\(E\quad \beta=0.820\quad \theta_{\mathrm{d}}=-65.0^\circ\)}
	\end{subfigure}
	
	\medskip
	
	\begin{subfigure}[t]{0.45\textwidth}
		\centering
		\includegraphics[width=\linewidth]{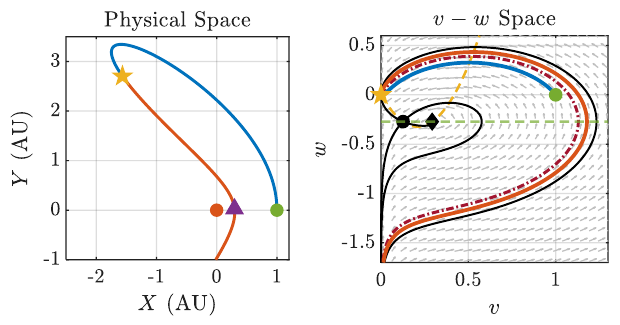}
		\caption*{\(C\quad \beta=0.613\quad \theta_{\mathrm{d}}=-26.3^\circ\)}
	\end{subfigure}
	\begin{subfigure}[t]{0.45\textwidth}
		\centering
		\includegraphics[width=\linewidth]{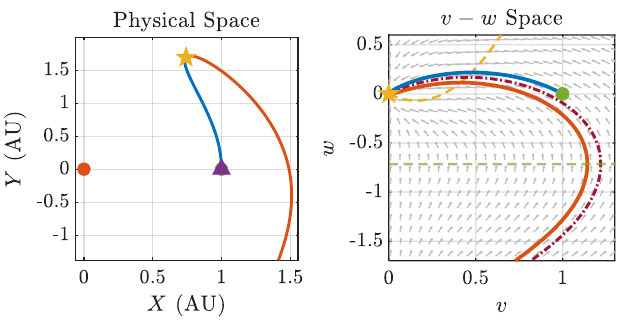}
		\caption*{\(F\quad \beta=0.874\quad \theta_{\mathrm{d}}=-55.0^\circ\)}
	\end{subfigure}
	
	\caption{Typical trajectories for the six parameter regions of the R-type SFDS in Fig.~\ref{fig:feasibility_r_type}.}
	\label{fig:typical_trajectories}
\end{figure*}

As illustrated in Fig.~\ref{fig:feasibility_analysis}, Regions 1–3 in the parameter space correspond to the scenarios where the initial point lies within the respective \textit{Region 1-3} of the $v$-$w$ space under the existence of equilibria, whereas Region 4 represents the no equilibrium point case. 
Additionally, "Int" (Internal) and "Ext" (External) designate whether the H-reversal trajectory turns inward or outward in physical space, with the self-returning trajectory serving as their boundary.
For the RS, the range of $\alpha$ is relatively restricted, with an H-reversal onset threshold of $\beta = 0.49$ ($\alpha = -23.7^\circ$).
In contrast, the R-type SFDS provides a significantly wider range of $\theta_{\text{d}}$ for H-reversal trajectory generation, lowering the minimum required $\beta$ to $\text{0.43}$ ($\theta_{\text{d}} = -68.2^\circ$).
Notably, the T-type SFDS is incapable of generating a self-returning trajectory, thereby completely precluding external H-reversal trajectory; its minimum generating threshold is limited to $\beta = \text{0.45}$ ($\theta_{\text{d}} = -90^\circ$). 
Figure~\ref{fig:typical_trajectories} presents the trajectories corresponding to the representative parameter combinations within the six regions in Fig.~\ref{fig:feasibility_r_type}.

\subsubsection{Constrained Perihelion Scenario}
For the R-type SFDS, numerical analysis confirms that $r_\text{p}$ decreases monotonically as $\theta_{\text{d}}$ becomes more negative. Thus, for a given $\beta$, the search starts from the $\theta_{\text{d}}^{\text{self-returning}}$, which is the boundary between the Int and Ext of Regions 3 and 4 in Fig.~\ref{fig:feasibility_r_type}. From this starting point, $\theta_{\text{d}}$ is progressively decreased until the specified $r_\text{p}$ constraint is satisfied.
This 
can be determined iteratively via a numerical root-finding algorithm as Algorithm~\ref{alg:SolveThetaD}.
where $\bm{X}_0$ denotes the initial state, $r_{\text{p,target}}$ represents the target perihelion distance, and $\epsilon_{\text{p}}$ is the tolerance of perihelion constraint, which is set to $10^{-8}$. $\theta_{\text{d,Low}}^{\text{H-rev}}$ and $\theta_{\text{d,Up}}^{\text{H-rev}}$ denote the lower and upper boundaries of the union of Regions 3 and 4 in Fig.~\ref{fig:feasibility_r_type}, thereby ensuring that the optimized $\theta_{\text{d}}^{\text{sol}}$ successfully yields an H-reversal trajectory. Similarly, the RS shares the same monotonicity and search strategy.
\begin{algorithm}[htbp]
	\caption{Perihelion-Constrained $\theta_{\text{d}}$ Solver}
	\label{alg:SolveThetaD}
	\begin{algorithmic}[1]
		\Function{SolveThetaD}{$t_0; \beta, r_{\text{p,target}}, \epsilon_{\text{p}}$}
		\State $\bm{X}_0 = [\bm{r}^{\top}_{\text{earth}}(t_0), \bm{v}^{\top}_{\text{earth}}(t_0)]^{\top}$
		\State $r_{\text{p}}(\theta_{\text{d}}) = \min_{t} r_{\text{SFDS}}(t; \bm{X}_0, \theta_{\text{d}}, \beta)$
		\State $\theta_{\text{d}}^{\text{sol}} \leftarrow \operatorname{rootfind}\left( e(\theta_{\text{d}})=r_{\text{p}}(\theta_{\text{d}})-r_{\text{p,target}}=0 \; \middle|\; \theta_{\text{d}} \in [\theta_{\text{d,Low}}^{\text{H-rev}}, \theta_{\text{d,Up}}^{\text{H-rev}}],\, \theta_{\text{d}}^{\text{self-returning}},\, \|e\| \le \epsilon_{\text{p}} \right)$
		\State \Return $\theta_{\text{d}}^{\text{sol}}$
		\EndFunction
	\end{algorithmic}
\end{algorithm}

Conversely, for the T-type SFDS, $r_\text{p}$ increases monotonically as $\theta_{\text{d}}$ decreases, prompting an initial evaluation at the extreme boundary of $\theta_{\text{d}} = -90^\circ$. If this case fails to produce a valid H-reversal trajectory or results in $r_\text{p} < r_\text{p,target}$, no feasible $\theta_{\text{d}}^{\text{sol}}$ exists for the given $\beta$. Otherwise, if $r_\text{p} > r_\text{p,target}$ at $\theta_{\text{d}} = -90^\circ$, $\theta_{\text{d}}$ is progressively increased until $r_\text{p} = r_\text{p,target}$, defining the upper bound of the feasible region. This also could be done via a root-finding algorithm.

\begin{figure*}[htbp]
	\centering
	\begin{subfigure}{0.26\textwidth}
		\centering
		\includegraphics[height=5cm]{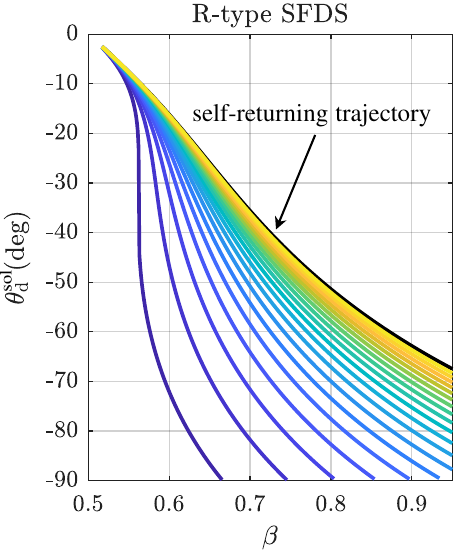}
		\caption{R-type SFDS}
		\label{fig:fea_cons_ds_r}
	\end{subfigure}
	\begin{subfigure}{0.26\textwidth}
		\centering
		\includegraphics[height=5cm]{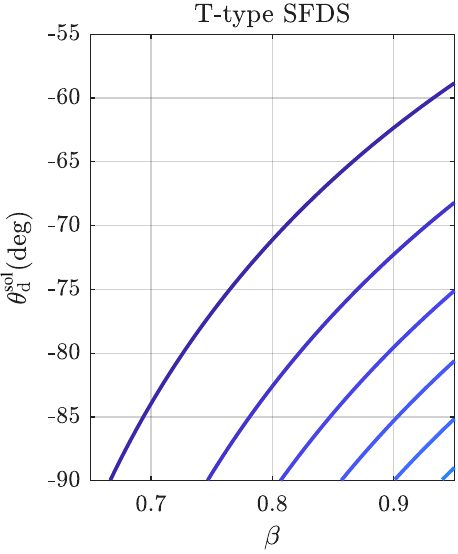}
		\caption{T-type SFDS}
		\label{fig:fea_cons_ds_t}
	\end{subfigure}
	\begin{subfigure}{0.26\textwidth}
		\centering
		\includegraphics[height=5cm]{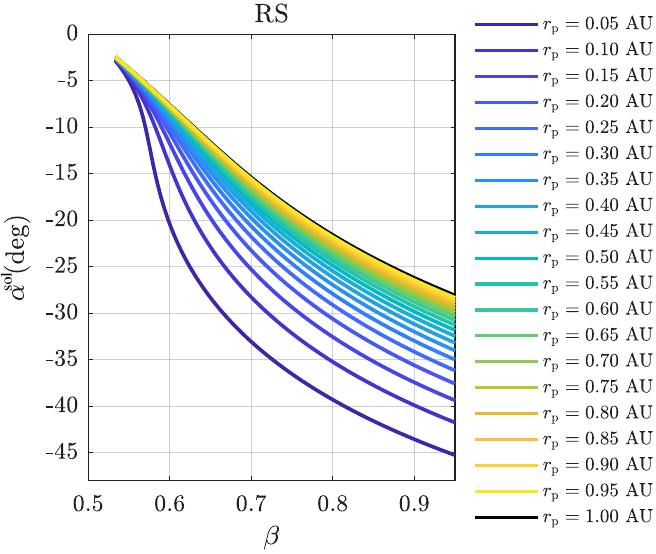}
		\caption{RS}
		\label{fig:fea_cons_ss}
	\end{subfigure}
	\caption{Changes in the boundaries of one-stage $\theta_{\text{\normalfont d}}$ feasibility regions under different perihelion constraints.}
	\label{fig:perihelion_constraints_comparison}
\end{figure*}
As illustrated in Fig.~\ref{fig:perihelion_constraints_comparison}, the lower limit of the parametric feasibility region for H-reversal orbits shrinks rapidly as the $r_{\text{p,target}}$  increase. 
Both the R-type SFDS and the RS exhibit a minimum $\beta$ converging toward nearly 0.52 to generate an H-reversal trajectory.
The $r_{\text p}$ contours become extremely dense as their values approach 1~AU, this problem is particularly prominent for RS. 
Consequently, an RS imposes stringent requirements on attitude stabilization and pointing accuracy. 
Conversely, an R-type SFDS demands only basic attitude maintenance,
since high-precision control of the diffraction angle can be achieved via electrically controlled LCPGs~\cite{ta_dynamic_2026}.
The T-type SFDS is incapable to satisfy $r_{\text p} \ge 0.3$ AU within the practical range of $\beta < 1$ owing to its limited normal force.
Consequently, the T-type configuration is excluded from subsequent analyses.

\subsection{Two-Stage $\theta_{\text{\normalfont d}}$ Strategy Feasibility Analysis}
\label{subsec:two_stage_feasibility}
As illustrated in Fig.~\ref{fig:perihelion_constraints_comparison}, under the one-stage $\theta_{\text d}$ strategy with a perihelion distance constraint, smaller $\beta$ inherently yield a smaller $\vert{}\theta_{\text d}\vert{}$. 
This significantly reduces the transverse thrust, particularly during the SPA phase, where the SRP cannot be fully utilized, degrading the efficiency of the H-reversal maneuver. 
The introduced two-stage $\theta_{\text d}$ control strategy could break through this bottleneck. 
A straightforward approach is to employ a smaller $\vert{}\theta_{\text d}\vert{}$ prior to the H-reversal point to secure the perihelion distance, and subsequently transition to a larger $\vert{}\theta_{\text d}\vert{}$ during the SPA acceleration phase to maximize the H-reversal efficiency.
The feasibility domain analysis is conducted below. 

\begin{figure*}[htbp]
	\centering
	\includegraphics[width=1\textwidth]{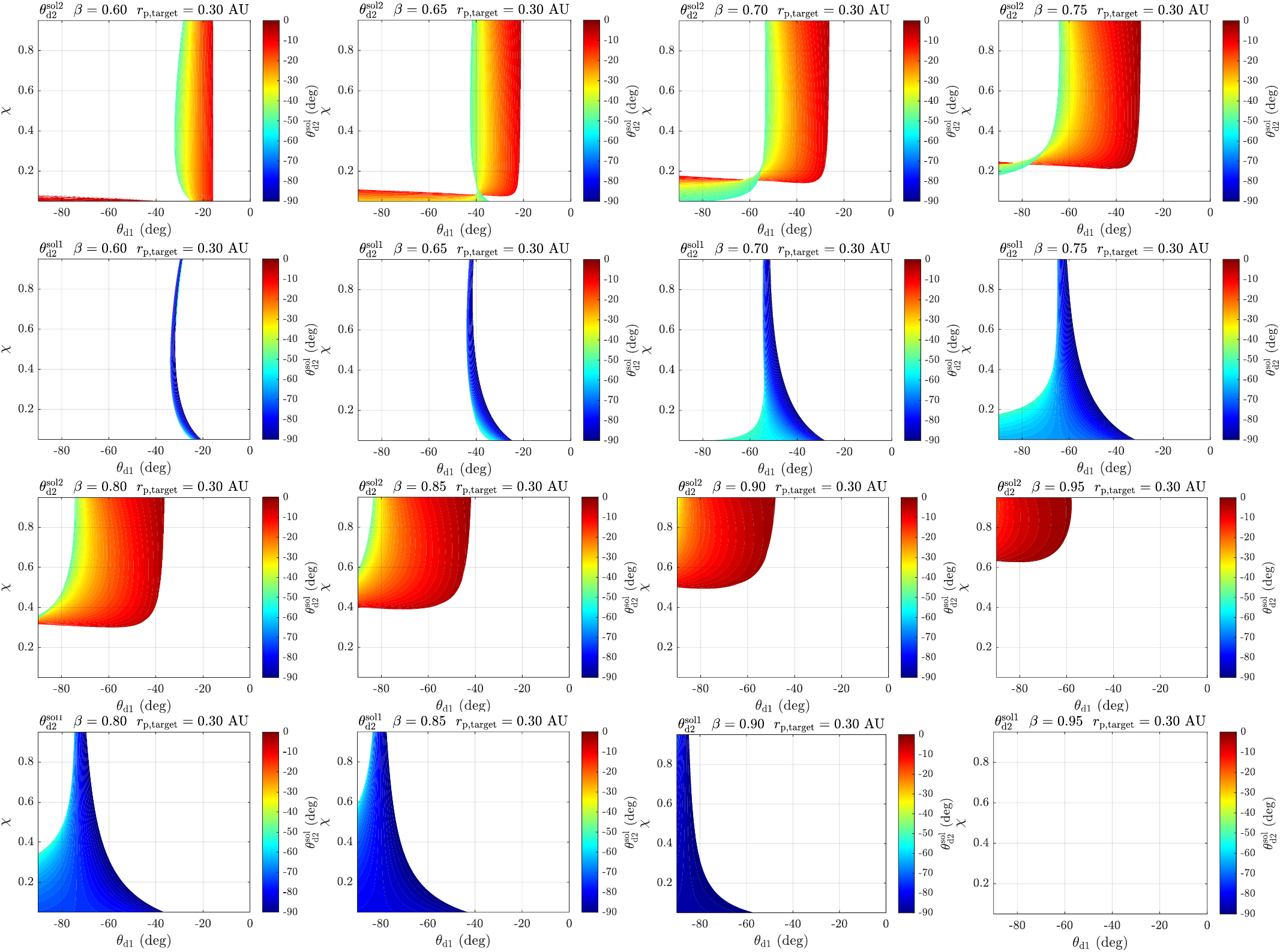}
	\caption{Two-stage $\theta_{\normalfont{\text{d}}}$ feasibility regions of R-type SFDS under different values of $\beta$ with $r_{{\normalfont{\text{p,target}}}}=0.3\text{ AU}$.}
	\label{fig:twofeasible}
\end{figure*}

For the convenience of analysis, a non-dimensional switching time parameter is defined as:
\begin{equation}
	\chi = \frac{t_{\text{sw}}-t_0}{t_{h}-t_0}
\end{equation}
where $t_{\text{sw}}$ represents the switching time and $t_{h}$ represents the epoch when $h=0$. 
The switching state $\boldsymbol{X}_{\text{sw}}$ is obtained via forward integration from $t_0$ to $t_{\text{sw}}$ under $\theta_{\text{d1}}$, 
and is subsequently mapped to the point $(v_{\text{sw}}, w_{\text{sw}})$ in the $v$-$w$ space.
Numerical experiments indicate that, there is no monotonic relationship between $\theta_{\text{d2}}$ and $r_{\text{p}}$ in this case.
To solve for the $\theta_{\text{d2}}$ that satisfies the perihelion constraint and ensure to form a retrograde orbit, the parameter $\theta_{\text{d2}}$ is uniformly discretized into $N = 100$ values within the interval $[-90^\circ, 0^\circ]$. 
Each grid point $\theta_{\text{d2}}^{k}$ along with $\beta$ uniquely determines a pair of parameters $(\xi_2, \eta_2)$ according to Eqs.~\eqref{eq:xi_eta_relation_with_beta_thetad}, thereby defining the phase space.
Because an H-reversal trajectory can only be formed when $(v_{\text{sw}}, w_{\text{sw}})$ lies within Region 3, 
candidate intervals $[\theta_{\text{d2}}^{k}, \theta_{\text{d2}}^{k+1}]$ are screened based on this criterion. 
If $r_{\text{p,target}}$ lies between $r_{\text{p}}(\theta_{\text{d2}}^{k})$ and $r_{\text{p}}(\theta_{\text{d2}}^{k+1})$, a root-finding algorithm is executed within that interval to precisely locate the exact solution $\theta_{\text{d2}}^{\text{sol}}$.
Numerical calculations indicate that there may be zero, one, or two solutions for $\theta_{\text{d2}}^{\text{sol}}$. Accordingly, solutions satisfying $\theta_{\text{d2}}^{\text{sol}} \le -50^\circ$ and $\theta_{\text{d2}}^{\text{sol}} > -50^\circ$ are designated as $\theta_{\text{d2}}^{\text{sol1}}$ and $\theta_{\text{d2}}^{\text{sol2}}$, respectively. 
This computational framework is formalized in Algorithm~\ref{alg:Algorithm2_optimized}.
The calculated feasible region across $\theta_{\text{d1}}$ from $-90^\circ$ to $0^\circ$ and $\chi$ from 0.05 to 0.95 with different $\beta$ under $r_{\text p} = 0.3$ AU is illustrated in Fig.~\ref{fig:twofeasible}.

\begin{figure*}[htbp]
	\centering
	\begin{subfigure}{0.4\textwidth}
		\centering
		\includegraphics[height=5.5cm]{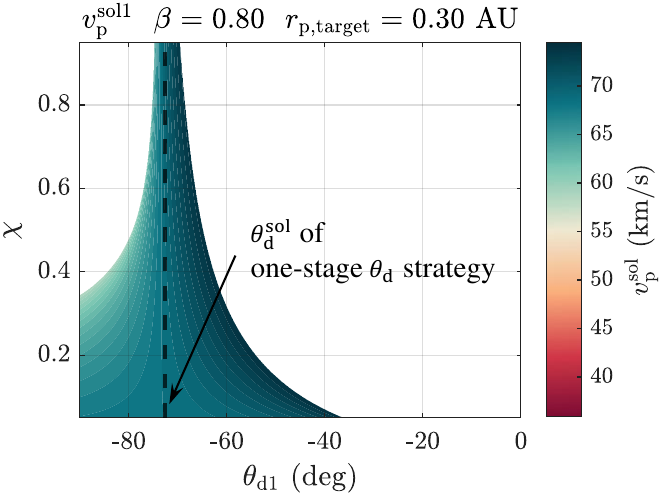}
		\caption{Perihelion-velocity of $\theta_{\normalfont{\text{d2}}}^{\normalfont{\text{sol1}}}$}
		\label{fig:perihelion_velocity_sol1}
	\end{subfigure}
	\hspace{1cm}
	\begin{subfigure}{0.4\textwidth}
		\centering
		\includegraphics[height=5.5cm]{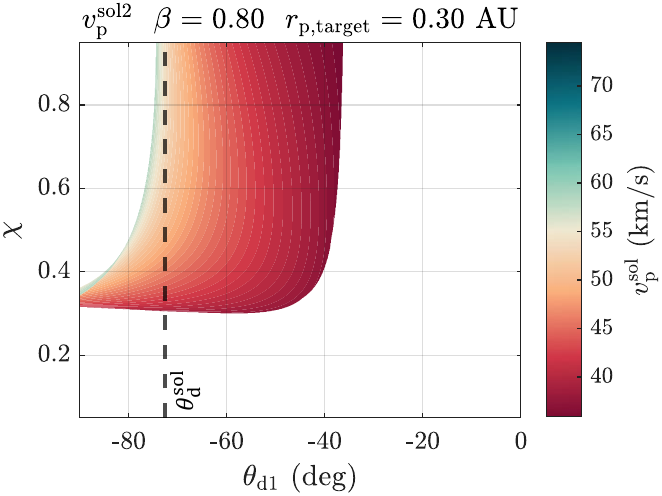}
		\caption{Perihelion-velocity of $\theta_{\normalfont{\text{d2}}}^{\normalfont{\text{sol2}}}$}
		\label{fig:perihelion_velocity_sol2}
	\end{subfigure}
	\caption{Perihelion-velocity distributions corresponding to the feasibility regions of two-stage $\theta_{\text{\normalfont d}}$ strategy under $\beta=0.8$ and $r_{\normalfont{\text{p,target}}}=0.3~\text{AU}$.}
	\label{fig:perihelion_velocity}
\end{figure*}

Figure~\ref{fig:perihelion_velocity} presents the perihelion velocity ($v_{\text{p}}^{\text{sol}}$)
corresponding to each feasible solution under $\beta=0.8$.
Comparison with Fig.~\ref{fig:twofeasible} indicates that the distribution of perihelion velocity is closely consistent with that of $\theta_{\text{d2}}^{\text{sol}}$, a more negative $\theta_{\text{d2}}^{\text{sol}}$ produces a higher $v_{\text{p}}^{\text{sol}}$.
This trend is consistently observed throughout all $\beta$ and
therefore provides a general design guideline. 
Accordingly, only the smaller solution $\theta_{\text{d2}}^{\text{sol1}} \leq -50^\circ$ is retained in the subsequent trajectory design, 
and it is expected that trajectories with $\theta_{\text{d2}}^{\text{sol1}}$ approaching $-90^\circ$ will yield higher impact velocities in asteroid impact missions.
Moreover, the solutions with higher $v_{\text{p}}^{\text{sol}}$ are concentrated near the right-hand boundary of the feasible domain. 
In the subsequent trajectory optimization, $\theta_{\text{d1}}$ can be initialized from the one-stage solution $\theta_{\text{d}}^{\text{sol}}$ and searched in the increasing direction.
In addition, the two-stage feasible domain becomes narrower as $\beta$ decreases,
therefore, restricting the initial guesses to the feasible domain and its vicinity can avoid the computational waste and improve convergence. 
The trajectory design method under the two-stage $\theta_{\text{d}}$ strategy is detailed in Section~\ref{subsection:Two-Stage Trajectory Design}.

\begin{algorithm}[hbt!]
	\caption{Perihelion-Constrained $\theta_{\text{d2}}$ Solver}
	\label{alg:Algorithm2_optimized}
	\begin{algorithmic}[1]
		\Function{SolveThetaD2}{$t_0, \chi, \theta_{\text{d1}}; \beta, r_{\text{p,target}}, \epsilon_{\text{p}}$}
		\State $\bm{X}_0 \leftarrow [\bm{r}_{\text{earth}}^{\top}(t_0), \bm{v}_{\text{earth}}^{\top}(t_0)]^{\top}$ 
		\State Integrate until $h = 0$ to find $t_{h}$, set $t_{\text{sw}} \leftarrow \chi t_h$ and evaluate $\bm{X}_{\text{sw}} \leftarrow \bm{X}(t_{\text{sw}})$
		\State Map $(v_{\text{sw}}, w_{\text{sw}}) \leftarrow \bm{X}_{\text{sw}}$
		\State Define grid $\Theta_{\text{d2}} = \{\theta_{\text{d2}}^{k}\}_{k=1}^N$ uniformly sampling $[-90^\circ, 0^\circ]$, initialize $\mathcal{S} \leftarrow \emptyset$
		
		\For{$k = 1$ \textbf{to} $N-1$} 
		\If{$(v_{\text{sw}}, w_{\text{sw}}) \in \text{Region 3}$ for both $\theta_{\text{d2}}^{k}$ and $\theta_{\text{d2}}^{k+1}$}
		\State Compute $e_{\text{p}}^{k} = r_{\text{p}}(\theta_{\text{d2}}^{k}) - r_{\text{p,target}}$ and $e_{\text{p}}^{k+1}$ via integration from $\bm{X}_{\text{sw}}$
		\If{$e_{\text{p}}^{k} \cdot e_{\text{p}}^{k+1} < 0$} 
		\State $\theta_{\text{d2}}^{\text{sol}} \leftarrow \operatorname{rootfind}\left( e_{\text{p}}(\theta_{\text{d2}})=0 \; \middle|\; \theta_{\text{d2}} \in [\theta_{\text{d2}}^{k}, \theta_{\text{d2}}^{k+1}],\frac{1}{2}(\theta_{\text{d2}}^{k} + \theta_{\text{d2}}^{k+1}),\, \|e_{\text{p}}\| \le \epsilon_{\text{p}} \right)$, $\mathcal{S} \cup \{\theta_{\text{d2}}^{\text{sol}}\}$ 
		\EndIf
		\EndIf
		\EndFor
		
		\State $\theta_{\text{d2}}^{\text{sol1}} \leftarrow \min\left(\left\{\theta_{\text{d2}}^{\text{sol}} \in \mathcal{S} \;\middle|\; \theta \le -50^\circ\right\}\right)$, $\theta_{\text{d2}}^{\text{sol2}} \leftarrow \max\left(\left\{\theta_{\text{d2}}^{\text{sol}} \in \mathcal{S} \;\middle|\; \theta > -50^\circ\right\}\right)$
		\State \Return $\theta_{\text{d2}}^{\text{sol1}}$, $\theta_{\text{d2}}^{\text{sol2}}$, $\bm{X}_{\text{sw}}$, $t_{\text{sw}}$
		\EndFunction
	\end{algorithmic}
\end{algorithm}

\section{H-reversal Trajectory Design on Asteroid Impact}

This section introduces the SFDS 2D H-reversal trajectories design methods for asteroid kinetic impact missions under both one-stage $\theta_{\text{d}}$ and two-stage $\theta_{\text{d}}$ strategies. 
To satisfy the geometric requirements of kinetic impact, maintain thermal safety, and ensure the formation of the H-reversal trajectory, the following constraints must be satisfied:
\begin{subequations}
\label{eq:constraints}
\begin{align}
	c_1 = r_{\text{SFDS}}(t_{\text{f}}) - r_{\text{ast}}(t_{\text{f}}) = 0, \label{eq:constraint1}\\
	c_2 = \theta_{\text{SFDS}}(t_{\text{f}}) - \theta_{\text{ast}}(t_{\text{f}}) = 0, \label{eq:constraint2}\\
	c_3 = r_{\text p} - r_{\text{p,target}} = 0,		\label{eq:constraint3}\\	
	g_1 = v_t(t_{\text{f}}) \le 0,	\label{eq:constraint4}\\
	g_2 = -v_r(t_{\text{f}}) \le 0,	\label{eq:constraint5}
\end{align}
\end{subequations}
where Eqs.~\eqref{eq:constraint1} and \eqref{eq:constraint2} define the terminal position equality constraints, ensuring zero position residual at the impact epoch $t_{\text{f}}$. 
H-reversal trajectories rely on a deep solar plunge to leverage the $1/r^2$ scaling of SRP,
due to the significant influence of perihelion distance on the objective function, equality constraint Eq.~\eqref{eq:constraint3} is directly used.
Furthermore, inequality constraints $g_1$ and $g_2$ regulate the motion state at $t_{\text{f}}$: Eq.~\eqref{eq:constraint4} guarantees that the spacecraft impacts the target in a retrograde direction,
whereas Eq.~\eqref{eq:constraint5} ensures that the perihelion has already been passed and the sail is flying outward after completing the SPA phase.

\subsection{One-Stage $\theta_{\text{\normalfont d}}$ Strategy Trajectory Design}

In this subsection, 
the problem is simplified into a search problem optimizing the launch epoch for terminal impact, to achieve this, a Perihelion-Constrained Phase-Matching Algorithm is proposed. 
Given a specific departure time $t_0$, at which the SFDS inherits the position and velocity of the Earth, Algorithm~\ref{alg:SolveThetaD} is invoked to determine $\theta_\text{d}$ that satisfies the perihelion constraint.
\begin{equation}
	\theta_{\text{d}}^{\text{sol}} \gets \Call{SolveThetaD}{t_0; \beta, r_{\text{p,target}}, \epsilon_{\text{p}}}.
\end{equation}
Concurrently, the trajectory is guaranteed to be an H-reversal trajectory, thereby satisfying Eq.~\eqref{eq:constraint3} and Eq.~\eqref{eq:constraint4}.
With the determination of $\theta_\text{d}^{\text{sol}}$, the equations of motion are integrated forward past the perihelion until the trajectory intersects the asteroid's orbit,
where the corresponding intersection epoch $t_{\text{f}}$ is recorded.
This process satisfies the post-perihelion outward flight requirement of Eq.~\eqref{eq:constraint5} and the radial distance equality constraint of Eq.~\eqref{eq:constraint1}.
The phase difference between the SFDS and the asteroid at $t_{\text f}$ can be expressed as a univariate function of the departure time $t_0$, denoted as $e_{\theta}(t_0) = \theta_{\text{SFDS}}(t_{\text{f}}) - \theta_{\text{ast}}(t_{\text{f}})$.
Therefore, the decision space is defined by a single variable $t_0$. 
The execution steps are outlined below.

First, $N$ departure times $t_0^{k}$ are uniformly sampled within the given search interval $[t_0^{\text{min}}, t_0^{\text{max}}]$.
By traversing the $N$ discrete grid points, $m$ intervals where $e_{\theta}(t_0^{k})$ changes sign are identified.
Subsequently, a root-finding algorithm is applied within these $m$ sign-changing intervals to solve for the impact departure time $t_0^m$ that satisfies $e_{\theta}(t_0^m) = 0$, thereby fulfilling the phase requirement of Eq.~\eqref{eq:constraint1}. 
The relative impact velocity is then calculated as $v_{\text{rel}}^m = \|\bm{v}_{\text{SFDS}}(t_{\text{f}}^{m}) - \bm{v}_{\text{ast}}(t_{\text{f}}^{m})\|$.
Finally, the optimal solution $v_{\text{rel}}^\text{opt}$ that maximizes the relative impact velocity is selected from the $m$ solutions.
The corresponding time of flight $\text{TOF}^{\text{opt}} = t_{\text{f}}^{\text{opt}} - t_0^{\text{opt}}$ are also recorded to represent the final impact performance of the SFDS under the given $\beta$ and $r_{\text{p,target}}$.
The complete execution procedure of the proposed algorithm is detailed in Algorithm \ref*{alg:Algorithm}. 
Additionally, for the fixed cone angle RS, this optimization procedure remains identical by substituting $\theta_{\text{d}}$ with $\alpha$.

\begin{algorithm}[htbp]
	\caption{Perihelion-Constrained Phase-Matching Algorithm}
	\label{alg:Algorithm}
	\begin{algorithmic}[1]
		\Require $t_0$, $\beta$, $r_{\text{p,target}}$, $\epsilon_{\text{p}}$, $\epsilon_{\text{f}}$
		\Ensure Optimal $t_{0}^\text{opt}, \theta_{\text{d}}^\text{opt}, v_{\text{rel}}^\text{opt}$
		\State Initialize: $t_0^{k} \in [t_0^{\text{min}}, t_0^{\text{max}}]$ for $k = 1, \dots, N$, $m \gets 0$, $\mathcal{S} \gets \emptyset$
		\For{$k = 1$ \textbf{to} $N-1$}
		\State $\theta_{\text{d}}^{k} \gets \Call{SolveThetaD}{t_0^{k}; \beta, r_{\text{p,target}}, \epsilon_{\text{p}}}$
		
		\State integrate with $\theta_{\text{d}}^{k}$ until $r_{\text{SFDS}}(t_{\text{f}}) = r_{\text{ast}}(\theta_{\text{SFDS}}(t_{\text{f}}^{k}))$, calculate $e_{\theta}^{k} = \theta_{\text{SFDS}}(t_{\text{f}}^{k}) - \theta_{\text{ast}}(t_{\text{f}}^{k})$
		\If {$e_{\theta}^{k} \cdot e_{\theta}^{k+1} < 0$} 
		\State $t_0^{m} \gets \text{rootfind}\left( e_{\theta}(t_0)=0 \;\middle|\; [t_0^{k},t_0^{k+1}], \frac{1}{2}(t_0^{k}+t_0^{k+1}), \|e_{\theta}\| \le \epsilon_{\text{f}} \right)$
		\State Integrate to calculate $t_{\text{f}}^{m}$, $\theta_{\text{d}}^{m}$ and $v_{\text{rel}}^{m} = \|\bm{v}_{\text{SFDS}}(t_{\text{f}}^{m}) - \bm{v}_{\text{ast}}(t_{\text{f}}^{m})\|$, $\mathcal{S} \gets \mathcal{S} \cup \left\{ t_0^{m}, \theta_{\text{d}}^{m}, v_{\text{rel}}^{m} \right\}$, $m \gets m+1$
		\EndIf
		\EndFor
		\State $m^* \gets \operatorname{argmax}_{m} \left\{v_{\text{rel}}^{m}\right\} $, $v_{\text{rel}}^\text{opt} \gets v_{\text{rel}}^{m^{*}}$, $t_{0}^\text{opt} \gets t_{0}^{m^{*}}$, $\theta_{\text{d}}^\text{opt} \gets \theta_{\text{d}}^{m^{*}}$ 
		\State \Return $t_{0}^\text{opt}, \theta_{\text{d}}^\text{opt}, v_{\text{rel}}^\text{opt}$
	\end{algorithmic}
\end{algorithm}

\subsection{Two-Stage $\theta_{\text{\normalfont d}}$ Strategy Trajectory Design}
\label{subsection:Two-Stage Trajectory Design}
Originally, the decision space is defined by a five-dimensional vector $\bm{Y}$:
\begin{equation}
	\bm{Y} = [t_0, \Delta t_1, \Delta t_2, \theta_{\text{d1}}, \theta_{\text{d2}}]^\top ,
\end{equation}
where $t_0$ is the epoch of Earth departure,  $\Delta t_1$ and $\Delta t_2$ represent the durations of the first and second stages, respectively, whereas $\theta_{\text{d1}}$ and $\theta_{\text{d2}}$ are the corresponding diffraction angles. The constraints are given in Eqs.~\eqref{eq:constraint1}--\eqref{eq:constraint5}.
The objective function is defined as:
\begin{equation}
	J(\bm{Y}) = -v_{\text{rel}}(t_f) = -\|\bm{v}_{\text{SFDS}}(t_f) - \bm{v}_{\text{ast}}(t_f)\| .
\end{equation}
To alleviate the computational difficulty of the optimization problem, the dimensionality and constraints are reduced by leveraging the functions and algorithms established in the previous section.
Specifically, given the $\beta$, $r_{\text{p,target}}$, $\theta_{\text{d1}}$ and $\chi$, the $\theta_{\text{d2}}$ that strictly satisfy the Eq.~\eqref{eq:constraint3} can be uniquely determined for any departure epoch $t_0$ via Algorithm~\ref{alg:Algorithm2_optimized}:
\begin{equation}	
	(\theta_{\text{d2}}^{\text{sol1}}, \theta_{\text{d2}}^{\text{sol2}}, \boldsymbol{X}_{\text{sw}}, t_{\text{sw}}) = \textsc{SolveThetaD2}(t_0, \chi, \theta_{\text{d1}}; \beta, r_{\text{p,target}}, \epsilon_{\text{p}}) .
\end{equation}
Since a diffraction angle with a larger absolute value yields a more efficient utilization of SRP, only the $\theta_{\text{d2}}^\text{{sol1}}$ is retained.
Notably, the algorithm inherently guarantees that the generated trajectory is an H-reversal trajectory through its hodograph characteristics, thereby naturally satisfying Eq.~\eqref{eq:constraint4}.
Similar to the one-stage, forward trajectory integration is performed from the switching state $\boldsymbol{X}_{\text{sw}}$ under $\theta_{\text{d2}}^\text{{sol1}}$.
This propagation continues until the SFDS passes perihelion and the trajectory intersects the asteroid's orbit, which defines the terminal epoch $t_{\text{f}}$. 
Consequently, Eq.~\eqref{eq:constraint5} and the radial distance requirement of Eq.~\eqref{eq:constraint1} are both satisfied.
The terminal phase difference $e_{\theta}$ between the SFDS and the asteroid is evaluated and expressed as a function of $t_0$, $\theta_{\text{d1}}$ and $\chi$, parameterized by the $\beta$ and $r_{\text{p,target}}$.
By embedding these constraints directly into the phase difference function, the terminal phase matching requirement of Eq.~\eqref{eq:constraint2} remains as the sole equality constraint in the optimization problem. 
As a result, the primary decision vector reduces to a three-dimensional space:
\begin{equation}
	\bm{Z} = [t_0, \chi, \theta_{\text{d1}}]^\top .
\end{equation}
This constraint-embedding mechanism effectively facilitates the numerical convergence of the non-linear optimization solver.
The two-stage $\theta_{\text{d}}$ trajectory design can be formulated as the following standard NLP problem:
\begin{equation}
	\min_{\bm{Z}_{\text{lb}} \le \bm{Z} \le \bm{Z}_{\text{ub}}} J(\bm{Z}) ~~\text{subject to}~~ e_{\theta}(\bm{Z}) = 0,
	\label{equ:optiprob}
\end{equation}
where $\bm{Z}_{\text{lb}}$ and $\bm{Z}_{\text{ub}}$ are the variable boundaries. 
To solve the problem, a Sequential Quadratic Programming (SQP) optimization method is employed. The overall procedure is outlined in Algorithm~\ref{alg:Algorithm2_TwoStage}.

\begin{algorithm}[htbp]
	\caption{Two-Stage $\theta_{\text{d}}$ Optimization Algorithm}
	\label{alg:Algorithm2_TwoStage}
	\begin{algorithmic}[1]
		\Require Initial guess $\bm{Z}^{(0)}$, boundaries $\bm{Z}_{\text{lb}}, \bm{Z}_{\text{ub}}$, parameters $\beta, r_{\text{p,target}}, \epsilon_{\text{p}}, \epsilon_{\text{f}}$
		\Ensure Optimal $\bm{Z}^\text{opt}, \theta_{\text{d2}}^\text{opt}, v_{\text{rel}}^\text{opt}$
		
		\State \label{line:init} Initialize: $\bm{Z} \gets \bm{Z}^{(0)}$
		\While {$\|e_{\theta}\| > \epsilon_{\text{f}}$ \textbf{or} not optimal}
		\State $(\theta_{\text{d2}}^{\text{sol1}}, \theta_{\text{d2}}^{\text{sol2}}, \bm{X}_{\text{sw}}, t_{\text{sw}}) \gets \textsc{SolveThetaD2}(\bm{Z}; \beta, r_{\text{p,target}}, \epsilon_{\text{p}})$
		
		\State Integrate from $\bm{X}_{\text{sw}}$ with $\theta_{\text{d2}}^{\text{sol1}}$ until $r_{\text{SFDS}}(t_{\text{f}}) = r_{\text{ast}}(\theta_{\text{SFDS}}(t_\text{f}))$ to get $t_{\text{f}}$
		\State $e_{\theta}(\bm{Z}) \gets \theta_{\text{SFDS}}(t_{\text{f}}) - \theta_{\text{ast}}(t_{\text{f}})$, and $J(\bm{Z}) \gets -v_{\text{rel}}(t_{\text{f}}) = -\|\bm{v}_{\text{SFDS}}(t_f) - \bm{v}_{\text{ast}}(t_f)\|$
		
		\State Update $\bm{Z}$ using SQP step to solve:
		\Statex \qquad\qquad $\min_{\bm{Z}} J(\bm{Z}) \quad \text{subject to} \quad e_{\theta}(\bm{Z}) = 0, \;\bm{Z}_{\text{lb}} \le \bm{Z} \le \bm{Z}_{\text{ub}}$
		\EndWhile
		
		\State $\bm{Z}^\text{opt} \gets \bm{Z}$, $\theta_{\text{d2}}^\text{opt} \gets \theta_{\text{d2}}$, $v_{\text{rel}}^\text{opt} \gets -J(\bm{Z})$
		\State \Return $\bm{Z}^\text{opt}, \theta_{\text{d2}}^\text{opt}, v_{\text{rel}}^\text{opt}$
	\end{algorithmic}
\end{algorithm}

\begin{figure*}[htbp]
	\centering
	\includegraphics[width=0.80\textwidth]{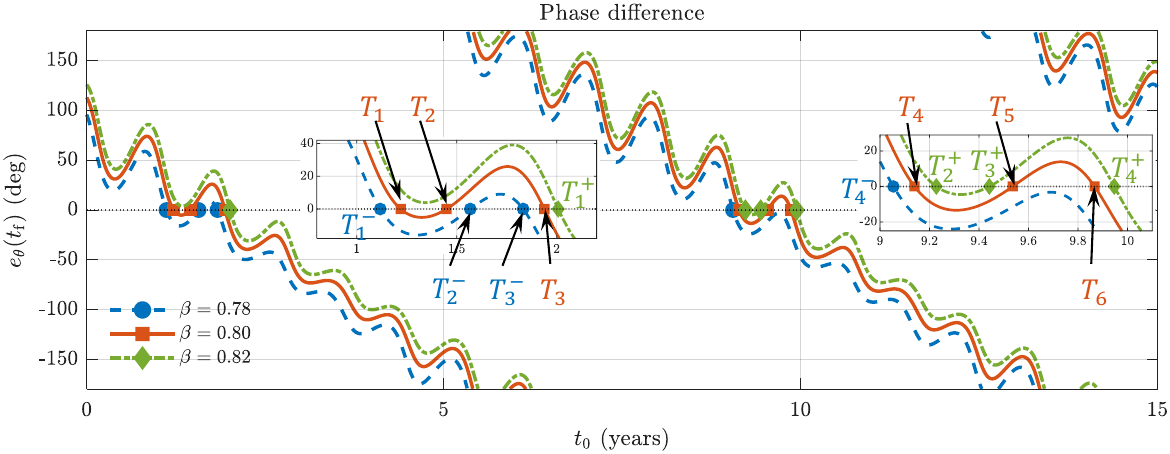}
	\caption{Phase-difference curves for
		$\beta=0.78$, $0.80$, and $0.82$.
		The colored markers indicate the launch epochs of 
		the solutions under one-stage $\theta_{\normalfont{\text{d}}}$ strategy.}
	\label{fig:t0_initial_channels}
\end{figure*}

The initialization procedure in Line~\ref{line:init} is described below. 
A tailored warm-start strategy is constructed using the solutions of the one-stage $\theta_{\text{d}}$ strategy and the feasible regions of two-stage $\theta_{\text{d}}$ strategy.
The treatment of the launch epoch $t_0$ is presented first.
To avoid missing feasible launch branches that may disappear as $\beta$ varies, as shown in Fig.~\ref{fig:t0_initial_channels}, 
the neighboring parameter set is defined as:
\begin{equation}
	\mathcal{B}_{\beta}
	=
	\left\{
	\beta_i:\left|\beta_i-\beta\right|\leq 0.02
	\right\}.
	\label{eq:neighboring_beta_set}
\end{equation}
For each $\beta_i\in\mathcal{B}_{\beta}$, the set of launch epochs for all solutions under one-stage $\theta_{\text{d}}$ strategy is retrieved:
\begin{equation}
	\mathcal{C}_{\beta_i} =	\left\{{t}_{0,\beta_i}^{\text{sol,}j}\right\}_{j=1}^{n_i},
\end{equation}
where ${t}_{0,\beta_i}^{\text{sol,}j}$ denotes the launch epoch of $j$th solution at $\beta_i$.
Each solution is expanded by one year in both directions:
\begin{equation}
	I_{i,j} = \left[ {t}_{0,\beta_i}^{\text{sol,}j} - \Delta t_{\mathrm w},	{t}_{0,\beta_i}^{\text{sol,}j} + \Delta t_{\mathrm w} \right],
	\qquad	
	\Delta t_{\mathrm w}=365.25~\mathrm{d}.
	\label{eq:t0_channel_expansion}
\end{equation}
Overlapping intervals are subsequently merged, yielding the final initialization interval for $t_0$:
\begin{equation}
	\bigcup_{\beta_i\in\mathcal{B}_{\beta}}
	\ \bigcup_{j=1}^{n_i} I_{i,j}
	= \bigcup_{m=1}^{N_{\mathrm w}} W_m,
	\qquad
	W_m\cap W_n=\varnothing ~~\text{for}~~ m \neq n,
	\label{eq:t0_search_domain}
\end{equation}
where $N_{\mathrm w}$ is the number of disjoint intervals after merging.
Taking $\beta = 0.80$ as a representative case, Fig.~\ref{fig:t0_initial_channels} and Table~\ref{tab:t0_launch_channels} demonstrate the aforementioned process.
During initialization, \(N_{t,\mathrm{init}}\) launch epoch guesses are uniformly distributed within each \(W_m\).

\begin{table*}[htbp]
	\centering
	\caption{Launch epochs of the solutions under one-stage $\theta_{\normalfont{\text{d}}}$ strategy
		and merged intervals used to initialize $t_0$.}
	\label{tab:t0_launch_channels}
	\small
	\setlength{\tabcolsep}{6pt}
	\begin{tabular}{ccccc}
		\toprule
		$\beta_i$ & Point & ${t}_{0,\beta_i}^{\text{sol,}j}$ ~(d) & $I_{i,j}$ ~(d) & Merged Interval\\
		\midrule
		
		0.78 & $T_1^-$ & 408.18  & $[42.93,\,773.43]$    & $W_1$ \\
		0.78 & $T_2^-$ & 572.64  & $[207.39,\,937.89]$   & $W_1$ \\
		0.78 & $T_3^-$ & 669.19  & $[303.94,\,1034.44]$  & $W_1$ \\
		0.78 & $T_4^-$ & 3306.92 & $[2941.67,\,3672.17]$ & $W_2$ \\
		\midrule
		\addlinespace[2pt]
		0.80 & $T_1$ & 445.47  & $[80.22,\,810.72]$    & $W_1$ \\
		0.80 & $T_2$ & 529.24  & $[163.99,\,894.49]$   & $W_1$ \\
		0.80 & $T_3$ & 708.00  & $[342.75,\,1073.25]$  & $W_1$ \\
		0.80 & $T_4$ & 3338.08 & $[2972.83,\,3703.33]$ & $W_2$ \\
		0.80 & $T_5$ & 3483.10 & $[3117.85,\,3848.35]$ & $W_2$ \\
		0.80 & $T_6$ & 3603.83 & $[3238.58,\,3969.08]$ & $W_2$ \\
		\midrule
		\addlinespace[2pt]
		0.82 & $T_1^+$ & 733.30  & $[368.05,\,1098.55]$  & $W_1$ \\
		0.82 & $T_2^+$ & 3370.28 & $[3005.03,\,3735.53]$ & $W_2$ \\
		0.82 & $T_3^+$ & 3448.52 & $[3083.27,\,3813.77]$ & $W_2$ \\
		0.82 & $T_4^+$ & 3632.84 & $[3267.59,\,3998.09]$ & $W_2$ \\
		
		\midrule
		\multicolumn{3}{r}{$W_1$:}
		& $[42.93,\,1098.55]$ & \\
		\multicolumn{3}{r}{$W_2$:}
		& $[2941.67,\,3998.09]$ & \\
		\bottomrule
	\end{tabular}
\end{table*}

\begin{figure}[htbp]
	\centering
	\includegraphics[width=0.5\textwidth]{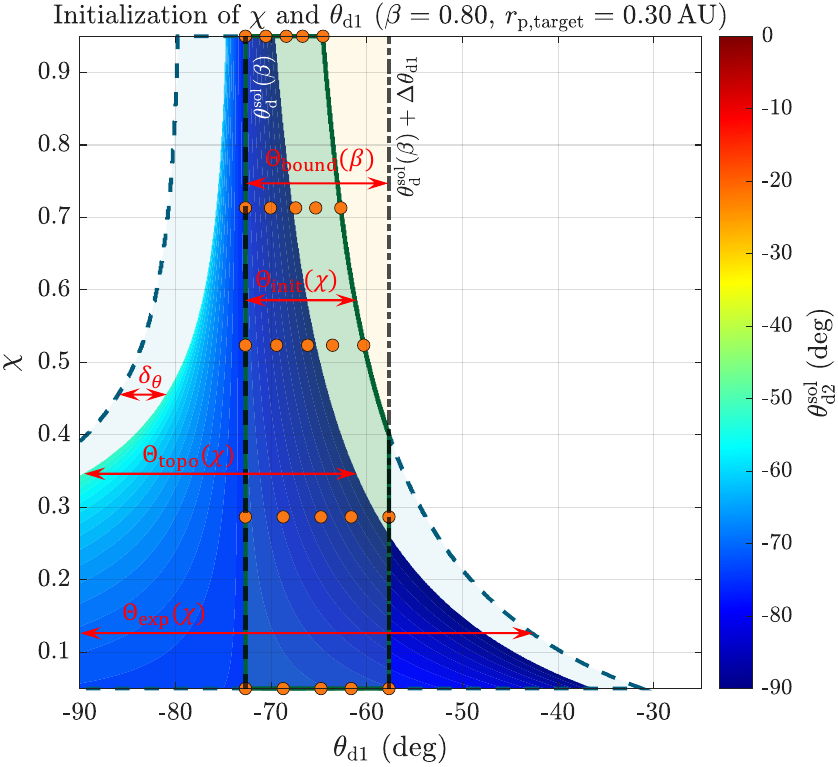}
	\caption{The tailoered initialization procedure and boundaries of $\theta_{\mathrm{d1}}$ and $\chi$ based on the feasible region.
		Only \(5\times5\) representative points are shown for clarity.}
	\label{fig:chi_th1_init}
\end{figure}

Next, the treatment of the remaining variables, $\chi$ and $\theta_{\mathrm{d1}}$, is presented. 
For a given $\beta$, the one-stage optimal diffraction angle $\theta_{\mathrm{d}}^{\mathrm{sol}}(\beta)$ is used as a reference. The bound set of $\theta_{\mathrm{d1}}$ is defined as
\begin{equation}
	\Theta_{\mathrm{bound}}(\beta) \triangleq [\theta_{\mathrm{d}}^{\mathrm{sol}}(\beta), \theta_{\mathrm{d}}^{\mathrm{sol}}(\beta)+\Delta\theta_{\mathrm{d1}}],
	\label{eq:theta1_bound_set}
\end{equation}
$\Delta\theta_{\mathrm{d1}}=15^\circ$ is adopted in this study. 
The bounds of $\chi$ are defined as $\chi_{\text{bound}}=[\chi_{\min},\chi_{\max}] = [0.05, 0.95]$.
The reason for these choices of boundaries has already been explained in Section~\ref{subsec:two_stage_feasibility}.
To concentrate the initial guesses near the feasible region, the precomputed two-stage feasibility topology for the corresponding $\beta$ is employed. 
This topology-guided initialization is inspired by the approach of Fu et al.~\cite{fu_energy_2026}.
Let $\Theta_{\mathrm{topo}}(\chi)$ denote the feasible set of $\theta_{\mathrm{d1}}$ at a given $\chi$. 
To account for possible boundary shifts between the circular orbit assumption and the actual elliptic orbit of the Earth, $\Theta_{\mathrm{topo}}(\chi)$ is expanded by a margin $\delta_\theta=5^\circ$, yielding the expanded feasible set
\begin{equation}
	\Theta_{\mathrm{exp}}(\chi) \triangleq \Theta_{\mathrm{topo}}(\chi) \oplus [-\delta_\theta,\delta_\theta],
	\label{eq:theta1_expanded_domain}
\end{equation}
where $\oplus$ denotes the Minkowski sum used for set expansion.
The initialization set for $\theta_{\mathrm{d1}}$ is expressed as
\begin{equation}
	\Theta_{\mathrm{init}}(\chi) = \Theta_{\mathrm{exp}}(\chi) \cap \Theta_{\mathrm{bound}}(\beta).
	\label{eq:theta1_initial_domain}
\end{equation}
During initialization, $N_{\chi,\mathrm{init}}$ values of $\chi$ are uniformly distributed over $\chi_{\text{bound}}$. 
For each $\chi_k^{(0)}$, $N_{\theta,\mathrm{init}}$ values of $\theta_{\mathrm{d1}}$ are then uniformly distributed within $\Theta_{\mathrm{init}}(\chi_k^{(0)})$. 
In this study, $N_{t,\mathrm{init}}=10$, $N_{\chi,\mathrm{init}}=N_{\theta,\mathrm{init}}=20$.
Finally, in each independent launch interval $W_m$, $N_{\mathrm{grid}}=N_{t,\mathrm{init}}N_{\chi,\mathrm{init}}N_{\theta,\mathrm{init}}=10\times20\times20=4000$ initial vectors $\bm{Z}^{(0)}$ are produced.
The above-mentioned initialization procedure is illustrated in Fig.~\ref{fig:chi_th1_init}.

To balance the trade-off between the terminal impact velocity and the mission duration, the objective function is reformulated as a weighted sum of the relative impact velocity and the time of flight ($\mathrm{TOF} = t_{\text{f}} - t_0$):
\begin{equation}
	L = q ( -v_{\text{rel}} ) + (1-q) \mathrm{TOF}, \quad q \in [0, 1]
	\label{eq:pareto_objective}
\end{equation}
where $q$ denotes the weighting coefficient, both $v_{\text{rel}}$ and $\mathrm{TOF}$ are normalized by the characteristic units defined in Section~\ref{sec:Dynamical Models}.
By varying the weighting coefficient $q$ to solve a series of optimization problems, the Pareto front can be mapped.
A homotopy approach is employed for the computation, where the optimization result from the previous weighting parameter serves as a warm start for the subsequent problem.

\section{Application in Apophis Deflection Mission}
In this section, the two strategies proposed in the preceding section for SFDS 2D H-reversal trajectory design are applied to the Apophis deflection mission. 
The effects of the sail lightness number~$\beta$ and the target perihelion distance~$r_\text{p,target}$ on the final terminal impact velocity and mission duration are evaluated.
\subsection{Apophis Information and experiment scenario}
The near-Earth asteroid 99942 Apophis has an equivalent diameter of approximately 340 m and an estimated mass of $4 \times 10^{10}$ kg. On April 13, 2029, it will perform a close Earth flyby at a nominal geocentric distance of approximately 38,000 km. This deep encounter will induce significant gravitational scattering, enlarging its orbital semi-major axis and transitioning its orbital classification from an Aten-type to an Apollo-type~\cite{li_apophis_2026}.

Rather than designing an individual, mission-specific trajectory, this section leverages the actual orbital dynamics of Apophis to demonstrate the superior time efficiency and impact kinetic energy of high-performance SFDS H-reversal trajectories.
In this study, to ensure a rigorous comparative analysis, we adopt a test scenario consistent with Gong et al.~\cite{gong_utilization_2011}, utilizing the initial orbital elements in the J2000 reference frame defined at the epoch of December 31, 2014, for numerical computations. 
The orbital elements used in this study are listed in Table~\ref{tab:orbital_elements}.
Given that the difference in orbital inclinations between the two bodies is relatively small, both are assumed to be 2D coplanar elliptic orbits within the ecliptic plane to simplify the analysis.
Consequently, the longitude of perihelion is defined as $\varpi=\Omega+\omega$, the orbital elements are converted to the planar polar state as:
\begin{equation}
	r=a(1-e\cos E),\qquad
	\theta=(\varpi+f),\qquad
	v_r=\frac{\mu e\sin f}{h},\qquad
	v_\theta=\frac{h}{r},\qquad
	h=\sqrt{\mu a(1-e^2)},
\end{equation}
the eccentric anomaly $E$ is first determined from the mean anomaly $M$ by solving Kepler's equation, after which the true anomaly $f$ is obtained from $E$.

To determine the optimal impact window, the search space for the initial launch epoch $t_0$ is swept over a 15-year interval from December 31, 2014 to December 31, 2029. This extensive range is designed to comprehensively encompass all typical relative geometric phases between Earth and Apophis.
The convergence tolerances for the terminal position and the perihelion constraint are $\epsilon_{\text{f}} = 10^{-8}\text{rad}$ and $\epsilon_{\text{p}} = 10^{-8}$ AU, respectively.

\begin{table}[htbp]
	\centering
	\caption{Initial classical orbital elements of Earth and Apophis at the epoch of December 31, 2014 (adapted from~\cite{gong_utilization_2011})}
	\label{tab:orbital_elements}
	\begin{tabular}{lccc}
		\hline\hline
		Orbital Element & Symbol & Earth & Apophis \\
		\hline
		Semi-major axis (AU) & $a$ & 1.00000 & 0.92239 \\
		Eccentricity & $e$ & 0.01672 & 0.19104 \\
		Inclination (rad) & $i$ & 0.00002 & 0.05814 \\
		Longitude of ascending node (rad) & $\Omega$ & 3.06142 & 3.56853 \\
		Argument of perihelion (rad) & $\omega$ & 5.01984 & 2.20548 \\
		Mean anomaly (rad) & $M_0$ & 6.23473 & 0.36147 \\
		\hline\hline
	\end{tabular}
\end{table}

\subsection{One-Stage $\theta_{\text{\normalfont d}}$ Strategy Numerical Results}

\begin{figure*}[htbp]
	\centering
	\begin{subfigure}{0.32\textwidth}
		\centering
		\includegraphics[height=4.5cm]{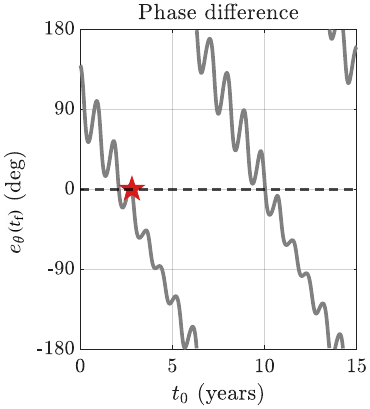}
		\caption{Phase difference}
		\label{fig:fix_phase}
	\end{subfigure}
	\hfill
	\begin{subfigure}{0.32\textwidth}
		\centering
		\includegraphics[height=4.5cm]{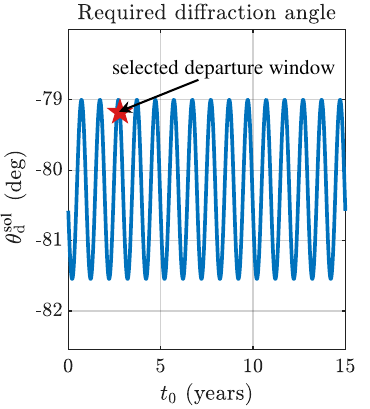}
		\caption{Required diffraction angle}
		\label{fig:fix_theta_d}
	\end{subfigure}
	\hfill
	\begin{subfigure}{0.32\textwidth}
		\centering
		\includegraphics[height=4.5cm]{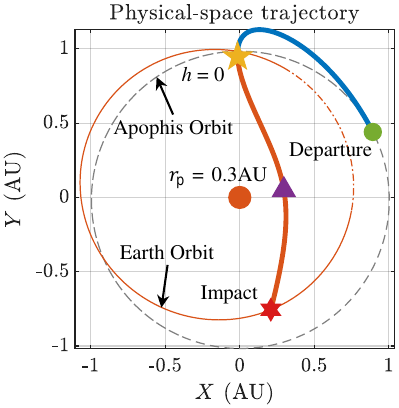}
		\caption{Physical-space trajectory}
		\label{fig:fix_trajectory}
	\end{subfigure}
	
	\begin{subfigure}{0.32\textwidth}
		\centering
		\includegraphics[height=4.5cm]{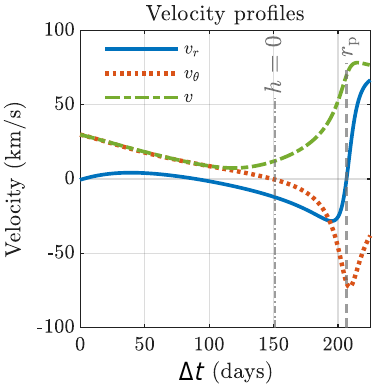}
		\caption{Velocity profiles}
		\label{fig:fix_velocity}
	\end{subfigure}
	\hfill
	\begin{subfigure}{0.32\textwidth}
		\centering
		\includegraphics[height=4.5cm]{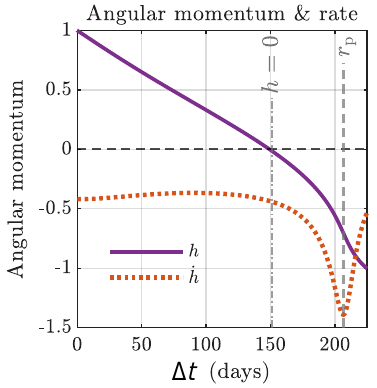}
		\caption{Angular momentum \& rate}
		\label{fig:fix_h}
	\end{subfigure}
	\hfill
	\begin{subfigure}{0.32\textwidth}
		\centering
		\includegraphics[height=4.5cm]{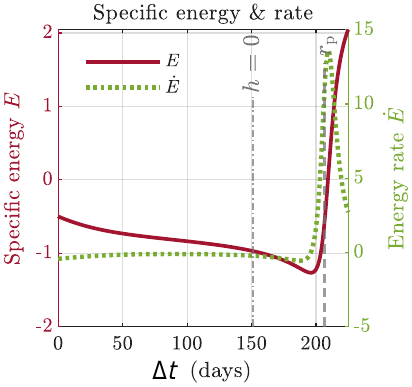}
		\caption{Specific energy \& rate}
		\label{fig:fix_E}
	\end{subfigure}
	
	\caption{One-stage strategy optimal solution of SFDS under parameter $\beta=0.85$ and $r_{\normalfont{\text{p,target}}}=0.3$ AU. }
	\label{fig:fix_angle_complete_analysis}
\end{figure*}

Based on the Algorithm~\ref{alg:Algorithm} presented in the preceding section, the 15-year candidate launch window is discrete into $N = 2000$ temporal grid points. 
Under the parameter of $\beta=0.85$ and $r_{\normalfont{\text{p,target}}}=0.3$ AU, Fig.~\ref{fig:fix_phase} illustrates the evolution of the terminal phase difference $e_{\theta}$ between SFDS and Apophis at the impact epoch as a function of the launch epoch $t_0$. 
The curve displays periodic fluctuations that reveals the synodic period of the Earth-Apophis system.
The zero-crossings of the phase curve represent the physically feasible launch windows where the terminal position constraint is satisfied. Among these candidate windows, the global optimal solution that maximizes the terminal relative impact velocity is highlighted by the red star.
Correspondingly, Fig.~\ref{fig:fix_theta_d} shows the required $\theta_{\text{d}}^{\text{sol}}$ distributed over the launch window to satisfy $r_{\text{p,target}} = 0.3$ AU. 
The subtle periodic oscillation of $\theta_{\text{d}}^{\text{sol}}$ between $-79^\circ$ and $-81.5^\circ$ is driven by the orbital eccentricity of the Earth.
The resulting optimal impact trajectory features a departure epoch of $t_0 = 1022.56$ days and a short flight duration of $\text{TOF} = 0.62$ years.
This rapid-response capability yields a hypervelocity terminal relative impact speed of $v_{\text{rel}}^{\text{opt}} = 102.46$ km/s. 
This velocity scale significantly exceeds the typical $10\text{--}15$ km/s range of traditional prograde interception missions and markedly surpasses the benchmark $60$ km/s retrograde interception limit established by conventional RS~\cite{mcinnes_deflection_2004}. 
The heliocentric physical-space trajectory is depicted in Fig.~\ref{fig:fix_trajectory}. 
The SFDS ultimately achieves a head-on collision with Apophis while the asteroid is moving toward its perihelion. 
To further demonstrate the state variations, Figs.~\ref{fig:fix_velocity}, \ref{fig:fix_h}, and \ref{fig:fix_E} outline the time-history profiles of the characteristic velocity components, specific orbital angular momentum, specific mechanical energy, and their respective derivatives. When approaching the perihelion, the extremely SRP leads to intense transverse acceleration, making the energy accumulation process highly localized within the SPA phase.

\subsubsection*{Comparison with RS}
\begin{figure*}[hbt!]
	\centering
	\begin{subfigure}[b]{0.32\textwidth}
		\centering
		\includegraphics[height=5.5cm]{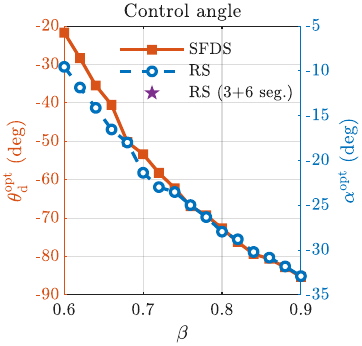}
		\caption{Control angle}
		\label{fig:sweep_beta1}
	\end{subfigure}
	\hfill
	\begin{subfigure}[b]{0.32\textwidth}
		\centering
		\includegraphics[height=5.5cm]{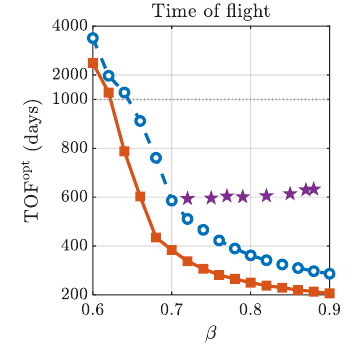}
		\caption{Time of flight}
		\label{fig:sweep_beta2}
	\end{subfigure}
	\hfill
	\begin{subfigure}[b]{0.32\textwidth}
		\centering
		\includegraphics[height=5.5cm]{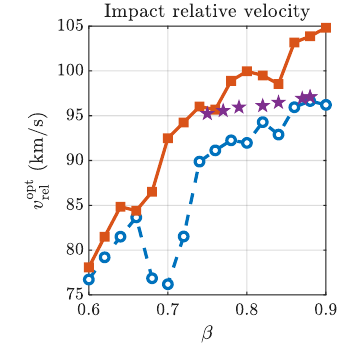}
		\caption{Impact relative velocity}
		\label{fig:sweep_beta3}
	\end{subfigure}
	
	\caption{Performance comparison between SFDS and RS under the one-stage strategy over a range of $\beta$ with $r_{\normalfont{\text{p,target}}} = 0.30$ AU).}
	\label{fig:sweep_beta}
\end{figure*}

To evaluate the influence of sail acceleration performance on the impact mission, a comparative analysis is performed between the SFDS and RS by varying the lightness number $\beta$ from 0.60 to 0.90 under a fixed target perihelion constraint of $r_{\text{p,target}} = 0.30$ AU. 
Fig.~\ref{fig:sweep_beta2} reveals the variation of the optimal flight time $\text{TOF}^{\text{opt}}$ with respect to $\beta$, indicating an exponential-like decay relationship. 
The flight time of the SFDS is shortened by a full 300 days compared to the RS at $\beta = 0.66$, the SFDS still reduces the flight time by nearly 200 days compared to the RS at $\beta = 0.70$. 
The evolution of the terminal relative velocity with respect to $\beta$ is illustrated in Fig.~\ref{fig:sweep_beta3}. Across all values of $\beta$, the impact velocity of the SFDS is generally 5--10 km/s higher than that of the RS. Particularly at $\beta = 0.70$, the impact velocity of the SFDS exceeds that of the RS by approximately 17 km/s. 
Notably, within the mission launch window of this study, the optimal trajectory for a smaller $\beta$ may execute the impact near the asteroid's perihelion, thereby yielding a higher relative velocity. 
Conversely, for a larger $\beta$, the impact may occur near the aphelion. 
This consequently leads to the localized non-monotonicity observed in Fig.~\ref{fig:sweep_beta3}.
The results from Gong et al.~\cite{gong_utilization_2011} are also plotted. It can be observed that even the one-stage $\theta_{\text{\normalfont d}}$ strategy of the SFDS outperforms the optimized results of the nine-stage piecewise attitude control strategy for the RS, which significantly simplifies the design of the attitude control system.

\begin{figure*}[hbt!]
	\centering
	\begin{subfigure}[b]{0.32\textwidth}
		\centering
		\includegraphics[height=5.5cm]{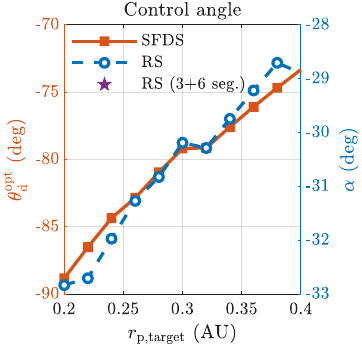}
		\caption{Control angle}
		\label{fig:sweep_rp1}
	\end{subfigure}
	\hfill
	\begin{subfigure}[b]{0.32\textwidth}
		\centering
		\includegraphics[height=5.5cm]{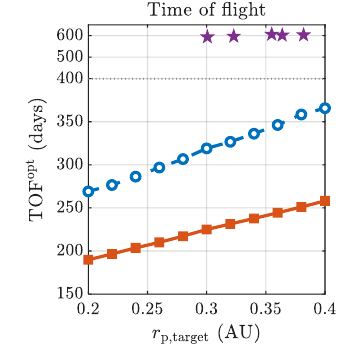}
		\caption{Time of flight}
		\label{fig:sweep_rp2}
	\end{subfigure}
	\hfill
	\begin{subfigure}[b]{0.32\textwidth}
		\centering
		\includegraphics[height=5.5cm]{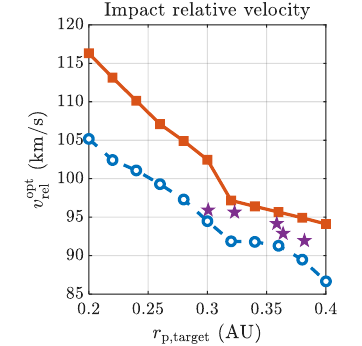}
		\caption{Impact relative velocity}
		\label{fig:sweep_rp3}
	\end{subfigure}
	
	\caption{Performance comparison between SFDS and RS under the one-stage strategy over a range of $r_{\normalfont{\text{p}}}$ with $\beta = 0.85$.}
	\label{fig:sweep_rp}
\end{figure*}

\begin{figure*}[hbt!]
	\centering
	\begin{subfigure}[b]{1\textwidth}
		\centering
		\includegraphics[width=1\textwidth]{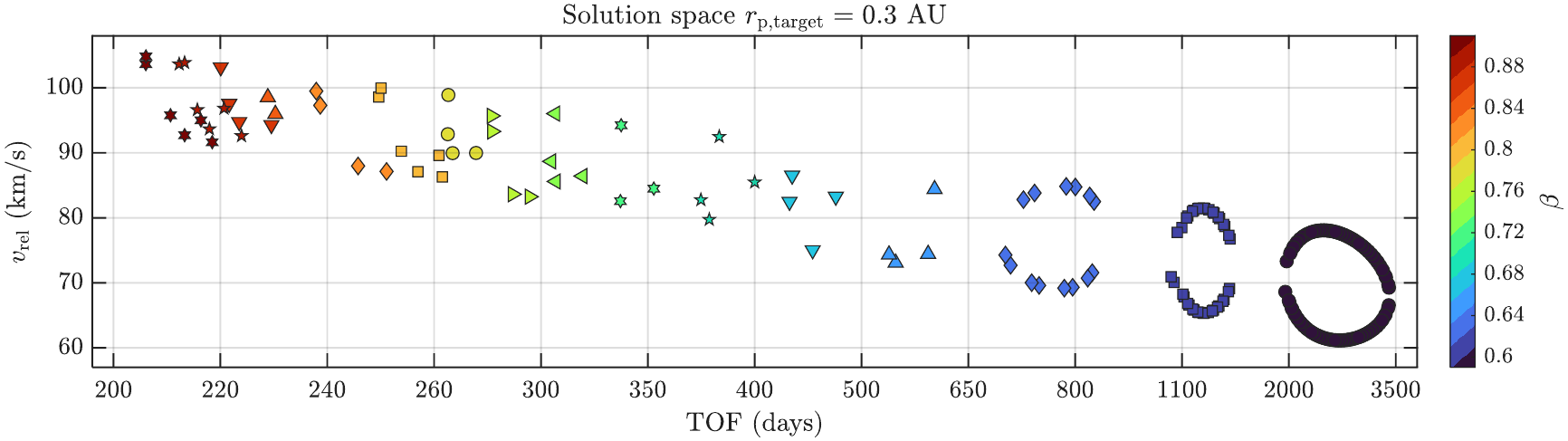} 
		\caption{Different $\beta$ with $r_{\normalfont{\text{p,target}}}=0.3\,\text{AU}$}
		\label{fig:solve_space_a}
	\end{subfigure}
	\\ \vspace{0.4cm} 
	\begin{subfigure}[b]{1\textwidth}
		\centering
		\includegraphics[width=1\textwidth]{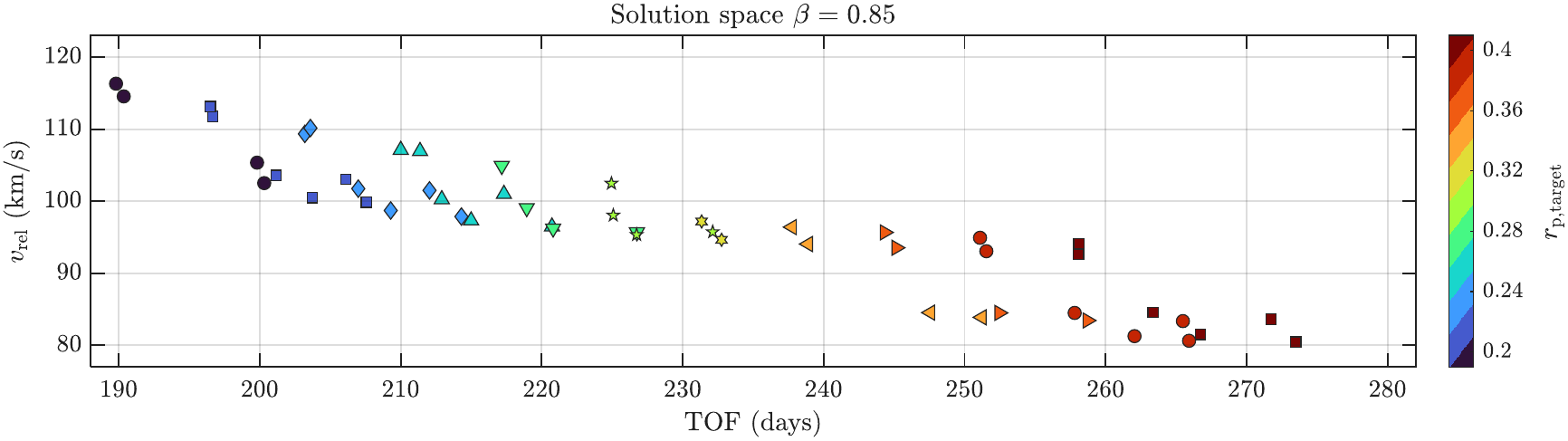} 
		\caption{Different $r_{\normalfont{\text{p,target}}}$ with $\beta=0.85$}
		\label{fig:solve_space_b}
	\end{subfigure}

	\caption{One-stage strategy solution space under different $\beta$ and $r_{\normalfont{\text{p,target}}}$}
	\label{fig:solve_space}
\end{figure*}

The constraint of the perihelion distance $r_{\text{p,target}}$ on the mission performance is investigated under $\beta = 0.85$.  
As shown in Fig.~\ref{fig:sweep_rp}, both the mission TOF and the impact velocities exhibit a linear relationship with the perihelion distance constraint.
As $r_{\text{p,target}}$ decreases from 0.4 AU to 0.2 AU, the mission duration of SFDS decreases from 260 days to 190 days, while the impact velocities  approaching 116 km/s.  
Therefore, the impact mission should minimize $r_{\text{p,target}}$ as much as possible to enhance mission performance. 
Likewise, across all perihelion cases, the SFDS outperforms the RS in terms of both flight time and impact velocity, demonstrating the prominent advantage of the SFDS in this mission scenario.
It is worth noting that Figs.~\ref{fig:sweep_beta} and \ref{fig:sweep_rp} only extract the solutions that maximize the relative impact velocity for each parameter configuration. However, as shown in Fig.~\ref{fig:fix_phase}, the evolution curve of the rendezvous phase difference with respect to the launch epoch often exhibits multiple zero-crossings. Therefore, Fig.~\ref{fig:solve_space} completely presents the solution space across all parameter configurations.

\subsection{Two-Stage $\theta_{\text{\normalfont d}}$ Strategy Numerical Results}

\begin{figure*}[hbt!]
	\centering
	\begin{subfigure}[b]{0.32\textwidth}
		\centering
		\includegraphics[width=\textwidth]{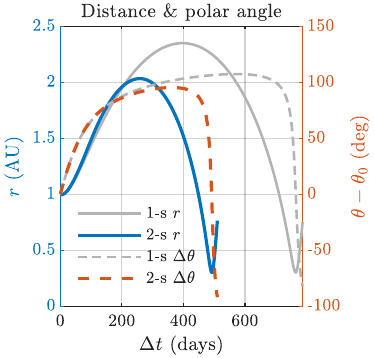} 
		\caption{Distance \& polar angle}
		\label{fig:sub_distance}
	\end{subfigure}
	\hfill
	\begin{subfigure}[b]{0.32\textwidth}
		\centering
		\includegraphics[width=\textwidth]{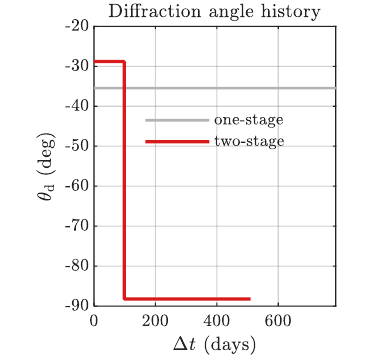} 
		\caption{Diffraction angle history}
		\label{fig:sub_diffraction}
	\end{subfigure}
	\hfill
	\begin{subfigure}[b]{0.32\textwidth}
		\centering
		\includegraphics[width=\textwidth]{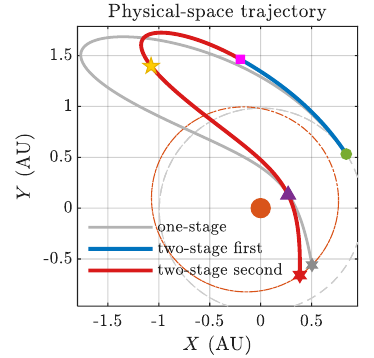} 
		\caption{Physical-space trajectory}
		\label{fig:sub_trajectory}
	\end{subfigure}
	\vspace{0.5cm} 
	\begin{subfigure}[b]{0.32\textwidth}
		\centering
		\includegraphics[width=\textwidth]{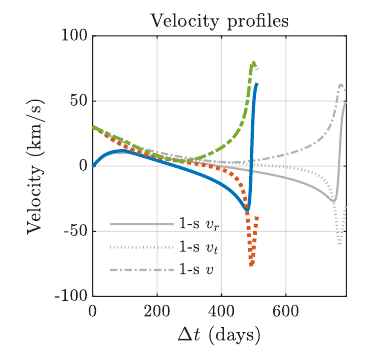} 
		\caption{Velocity profiles}
		\label{fig:sub_velocity}
	\end{subfigure}
	\hfill
	\begin{subfigure}[b]{0.32\textwidth}
		\centering
		\includegraphics[width=\textwidth]{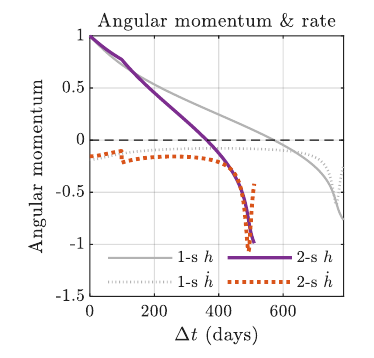} 
		\caption{Angular momentum \& rate}
		\label{fig:sub_momentum}
	\end{subfigure}
	\hfill
	\begin{subfigure}[b]{0.32\textwidth}
		\centering
		\includegraphics[width=\textwidth]{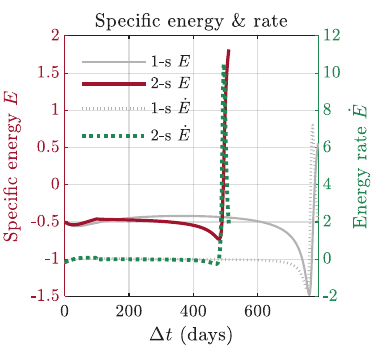} 
		\caption{Specific energy \& rate}
		\label{fig:sub_energy}
	\end{subfigure}
	\caption{Two-stage strategy optimal solution of SFDS under parameter $\beta=0.64$ and $r_{\normalfont{\text{p,target}}}=0.3$ AU. }
	\label{fig:twostagetraj}
\end{figure*}

The search boundaries for the design variables are set as $t_0 \in [0, 15 \times 365.25]\text{d}$, $\chi \in [0.05, 0.95]$ and $\theta_{\text{\normalfont d1}} \in \Theta_{\mathrm{bound}}(\beta)$. 
To facilitate numerical convergence, a tailored initialization strategy detailed in Section.~\ref{subsection:Two-Stage Trajectory Design} is implemented to generate the initial guess vectors $\bm{Z}^{(0)}$. 
To visualize the two-stage $\theta_{\text{d}}$ strategy of SFDS, a schematic diagram is illustrated in Fig.~\ref{fig:twostagetraj} under parameter $\beta=0.64$ and $r_{\normalfont{\text{p,target}}}=0.3$ AU.

\begin{figure*}[hbt!]
	\centering
	\begin{subfigure}[b]{0.33\textwidth}
		\centering
		\includegraphics[width=\textwidth]{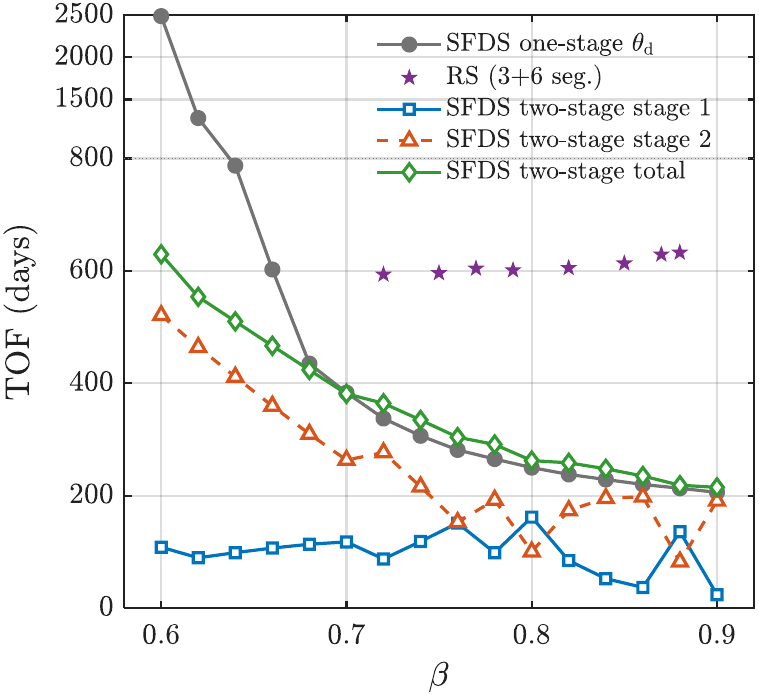}
		\caption{Time of flight}
		\label{fig:time_of_flight}
	\end{subfigure}
	\hfill 
	\begin{subfigure}[b]{0.33\textwidth}
		\centering
		\includegraphics[width=\textwidth]{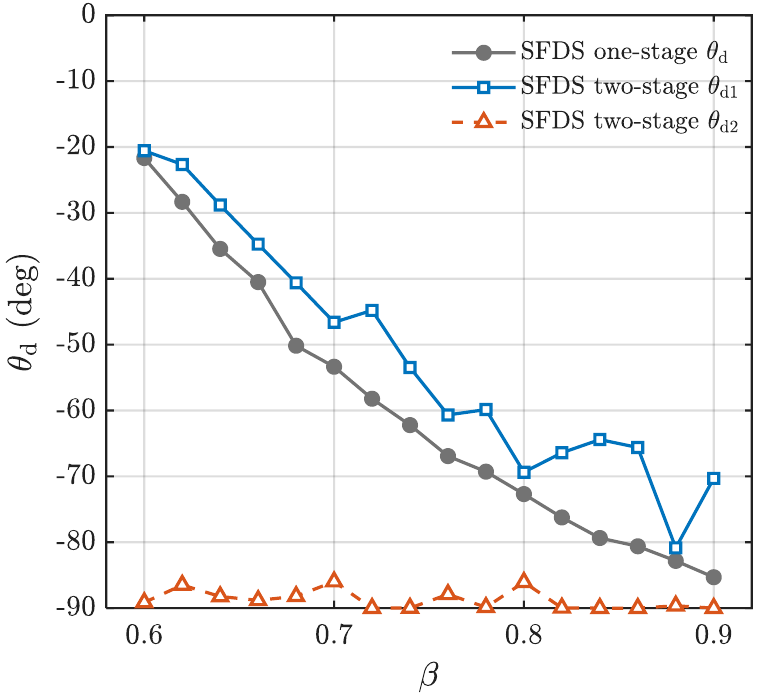}
		\caption{Diffraction angle}
	\label{fig:diffraction_angle}
	\end{subfigure}
	\hfill
	\begin{subfigure}[b]{0.33\textwidth}
		\centering
		\includegraphics[width=\textwidth]{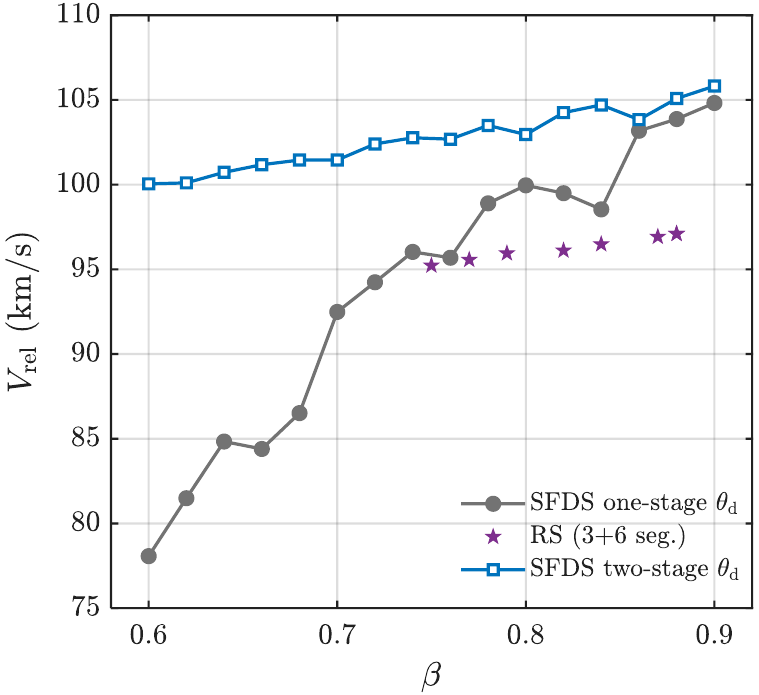}
		\caption{Impact relative velocity}
		\label{fig:impact_relative_velocity}
	\end{subfigure}
	\caption{Performance comparison of SFDS under the one-stage and two-stage $\theta_{\normalfont{\text{d}}}$ strategies over a range of $\beta$ with $r_{\normalfont{\text{p,target}}} = 0.30$ AU.}
	\label{fig:two_stage_sweep_beta}
\end{figure*}

The optimization results for different $\beta$ values are illustrated in Fig.~\ref{fig:two_stage_sweep_beta}. 
In the two-stage strategy, the second-stage $\theta_{\text{d2}}$ consistently approach -90°, which enhances the acceleration efficiency during the SPA phase. 
Consequently, compared with the traditional one-stage strategy, the two-stage strategy yields a systematically higher terminal relative impact velocity across all values of $\beta$. 
This performance improvement is particularly pronounced in the range of $\beta = 0.6$ to $0.75$.
Specifically, at $\beta = 0.62$, the two-stage strategy delivers a massive velocity increment of up to $19~\mathrm{km/s}$ while simultaneously compressing the total transfer duration by $318~\mathrm{days}$.
As $\beta$ increases, the advantage of the two-stage strategy gradually diminishes. 
This occurs because the $\theta_{\text{d}}^{\text{opt}}$ in the one-stage strategy progressively approaches $\theta_{\text{d1}}$ and $\theta_{\text{d2}}$ of the two-stage strategy.
Therefore, for practical missions, the two-stage strategy can be adopted at low $\beta$ to enhance performance, whereas the one-stage strategy can be chosen at high $\beta$ to simplify the diffractive microstructure design.
The results from Gong et al.~\cite{gong_utilization_2011} are also plotted in the figure. It can be observed that utilizing the SFDS substantially increases the impact velocity, reduces the mission duration, and simplifies the design of the attitude control system.

\begin{figure*}[hbt!]
	\centering
	\includegraphics[width=1\textwidth]{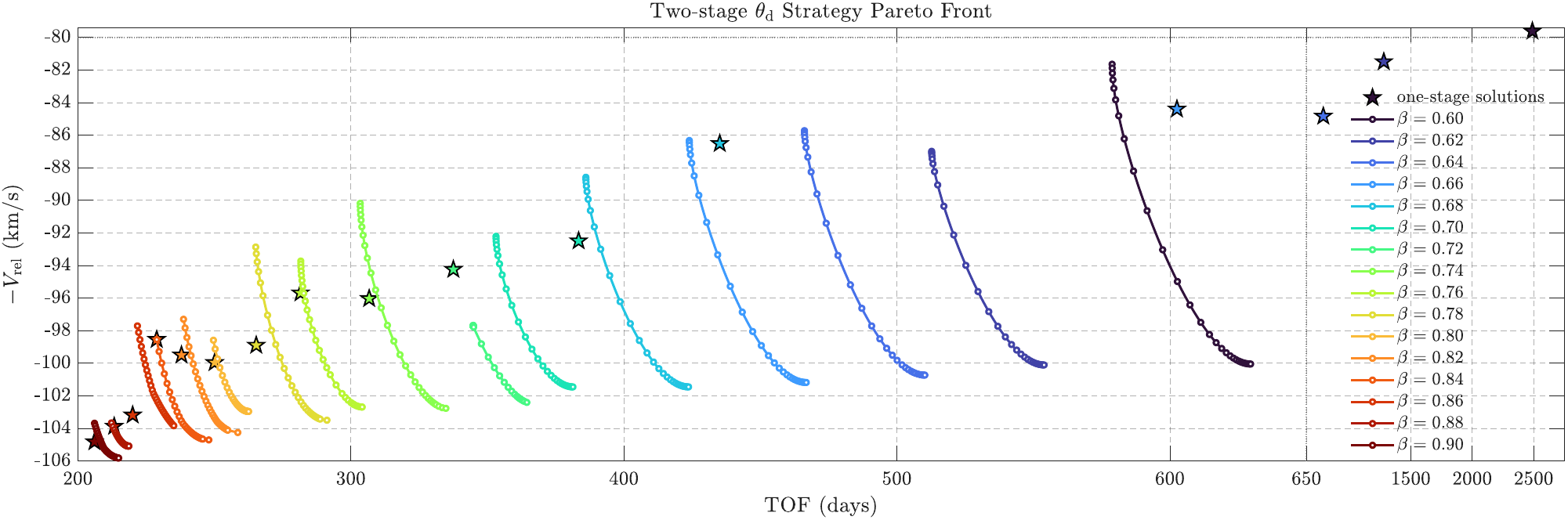}
	\caption{Pareto front of the SFDS under the two-stage $\theta_{\normalfont{\text{d}}}$ strategy over a range of $\beta$ with $r_{\normalfont{\text{p,target}}} = 0.30$ AU.}
	\label{fig:Pareto}
\end{figure*}

By varying the weights in the objective function~\eqref{eq:pareto_objective}, a series of optimal solutions is obtained, as shown in Fig.~\ref{fig:Pareto}. 
The Pareto fronts explicitly illustrate the inherent trade-off between the time of flight and the relative impact velocity. 
For an SFDS with a given $\beta$, achieving higher impact kinetic energy necessitates a longer TOF for orbital energy accumulation.
From a mission design perspective, the traditional one-stage strategy typically converges only to discrete design points (asterisk markers). 
In contrast, the proposed two-stage strategy yields continuous Pareto design curves. Based on the urgency of the launch window and the target asteroid's mass (i.e., the $v_{\text{rel}}$ requirement), designers can directly refer to these curves to select an appropriate $\beta$ and determine the required SFDS performance.
This advantage significantly expands the feasible design space, providing a more flexible decision-making framework for asteroid defense missions in emergency scenarios.

\section{Conclusion}

This paper investigated the theoretical feasibility and planetary defense applications of SFDS achieving H-reversal trajectories under both one-stage diffraction angle ($\theta_{\text{\normalfont d}}$) and two-stage $\theta_{\text{\normalfont d}}$ control strategies. 
By mapping the planar orbital dynamics into the velocity space, several key insights and methodological advancements have been established:
First, utilizing the hodograph method, the parametric feasibility domains and boundary characteristics for 2D H-reversal trajectories under perihelion distance constraint ($r_{\text{p,target}}$) were mapped for both strategies. 
The results show that, under the one-stage $\theta_{\text{\normalfont d}}$ strategy, the minimum lightness number ($\beta$) required for an R-type SFDS to achieve an H-reversal trajectory is approximately 0.52, 
whereas a T-type SFDS has difficulty achieving H-reversal trajectory when $r_{\text{p,target}}>0.3\mathrm{AU}$. 
For the two-stage $\theta_{\text{\normalfont d}}$ strategy, the feasible domain becomes progressively narrower and more elongated as $\beta$ decreases. 
Moreover, the distribution of the perihelion velocity closely aligned with the $\theta_{\text{\normalfont d2}}$ in the $\theta_{\text{\normalfont d1}}$--$\chi$ space, where $\chi$ denotes the switching time parameter.
The resulting feasibility domains provide methodological guidance for practical trajectory design of asteroid impact missions, while also offering a basis for generating suitable initial guesses for optimization.
For the one-stage $\theta_{\text{d}}$ strategy, a Perihelion-Constrained Phase-Matching algorithm was proposed to determine the launch epoch. 
For the two-stage $\theta_{\text{d}}$ strategy, an optimization framework was developed to reduce the high-dimensional, multi-constrained problem into a three-dimensional, single-constraint optimization. 
This reduction was achieved by embedding multiple constraints directly into the terminal phase requirement, while the initial guesses were efficiently guided by the previously mapped feasibility domain. 
Third, numerical simulations targeting Asteroid 99942 Apophis validated the advantages of the SFDS over the conventional RS. Under $\beta = 0.70$, the one-stage $\theta_{\text{d}}$ strategy increased the terminal relative velocity by 21\% and compressed the total mission duration by 35\% compared to the RS. 
The proposed two-stage $\theta_{\text{d}}$ strategy allows the SFDS to utilize the SRP more fully, further elevating the impact velocity to $100-106~\si{\kilo\meter\per\second}$. 
Additionally, a homotopy-driven Pareto front analysis was integrated to trace the trade-offs between flight time and impact velocity. 
In summary, this study establishes an analysis framework of SFDS H-reversal trajectory, establishing a feasible baseline for emergency asteroid kinetic impact missions.

\section*{Funding Sources}
The authors acknowledge the financial support from 
the National Natural Science Foundation of China (Grant No.12372044), 
the National Natural Science Foundation of China (Grant No. 12525204),
and
the Academic Excellence Foundation of BUAA for PhD Students.

\bibliography{references}
\end{document}